\documentclass[nofootinbib,twocolumn,superscriptaddress,longbibliography]{revtex4-1} 

\usepackage{latexsym}
\usepackage{graphics}
\usepackage{bm}
\usepackage[normalem]{ulem}
\usepackage{graphicx} 
\usepackage{color}
\usepackage{amsmath}
\usepackage{amssymb}
\usepackage{feynmp-auto}
\usepackage{slashed}
\usepackage{enumerate}
\usepackage{amsfonts, amssymb,amsmath,latexsym,graphics, graphicx,epsfig,multirow,comment,
,feyn,slashed,xcolor,afterpage, makecell} 
\usepackage{booktabs, blindtext}
\usepackage{tabularx,afterpage}
\usepackage[colorlinks=true
,urlcolor=blue
,anchorcolor=blue
,citecolor=blue
,filecolor=blue
,linkcolor=blue
,menucolor=blue
,linktocpage=true
,pdfproducer=medialab
,pdfa=true
]{hyperref}
\usepackage[utf8]{inputenc}
\usepackage{url}

\renewcommand{\arraystretch}{1.8}
\newcolumntype{L}[1]{>{\raggedright\let\newline\\\arraybackslash\hspace{0pt}}m{#1}}
\newcolumntype{C}[1]{>{\centering\let\newline\\\arraybackslash\hspace{0pt}}m{#1}}
\newcolumntype{R}[1]{>{\raggedleft\let\newline\\\arraybackslash\hspace{0pt}}m{#1}}

\newcommand {\be} {\begin {equation}}
\newcommand {\ee} {\end {equation}} 

\newcommand {\bes} {\begin {equation*}}
\newcommand {\ees} {\end {equation*}}

\newcommand{\beq}{\begin{equation}}
\newcommand{\eeq}{\end{equation}}
\newcommand{\bea}{\begin{eqnarray}}
\newcommand{\eea}{\end{eqnarray}}
\newcommand{\Eq}[1]{Eq.~(\ref{#1})}
\newcommand{\Eqs}[2]{Eqs.~(\ref{#1}) and (\ref{#2})}
\newcommand{\Sec}[1]{Sec.~\ref{#1}}

\newcommand{\Fig}[1]{Fig.~\ref{#1}}

\newcommand{\Ref}[1]{Ref.~\cite{#1}}
\newcommand{\Refs}[1]{Refs.~\cite{#1}}

\newcommand{\App}[1]{App.~\ref{#1}}

\newcommand{\Rsun}{R_{0}}
\newcommand{\zsun}{z_{0}}

\newcommand{\FdisE}{\ifmmode {\lvert F_{dis}(E_r,E_{int}) \rvert}^2 \else ${\lvert F_{dis}(E_r,E_{int}) \rvert}^2$\fi}
\newcommand{\Fmol}{\ifmmode {\lvert F_{mol}(\mathbf{q,\tilde{\mathbf{q}}}) \rvert}^2 \else ${\lvert F_{mol}(\mathbf{q},\tilde{\mathbf{q}}) \rvert}^2$\fi}

\newcommand{\Mfi}{\ifmmode {\lvert \mathcal{M}_{2-2} \rvert}^2 \else ${\lvert \mathcal{M}_{2-2} \rvert}^2$\fi}

\usepackage{csvsimple}

\begin{document}

\title{Testing Dark Matter and Modifications to Gravity\\ using Local Milky Way Observables}

\preprint{PUPT XXXX}

\author{Mariangela Lisanti}
\affiliation{Department of Physics, Princeton University, Princeton, NJ 08544}

\author{Matthew Moschella}
\affiliation{Department of Physics, Princeton University, Princeton, NJ 08544}

\author{Nadav Joseph Outmezguine}
\affiliation{Institute for Advanced Study, Princeton, NJ 08544}
\affiliation{Raymond and Beverly Sackler School of Physics and Astronomy, Tel-Aviv University, Tel-Aviv 69978, Israel}

\author{Oren Slone}
\affiliation{Princeton Center for Theoretical Science, Princeton University, Princeton, NJ 08544}

\date{\today}

\begin{abstract}
Galactic rotation curves are often considered the first robust evidence for the existence of dark matter. However, even in the presence of a dark matter halo, other galactic-scale observations, such as the Baryonic Tully-Fisher Relation and the Radial Acceleration Relation, remain challenging to explain. This has motivated long-distance, infrared modifications to gravity as an alternative to the dark matter hypothesis as well as various DM theories with similar phenomenology. In general, the standard lore has been that any model that reduces to the phenomenology of MOdified Newtonian Dynamics (MOND) on galactic scales explains essentially all galaxy-scale observables. We present a framework to test precisely this statement using local Milky Way observables, including the vertical acceleration field, the rotation curve, the baryonic surface density, and the stellar disk profile. We focus on models that predict scalar amplifications of gravity, \emph{i.e.},~models that increase the magnitude but do not change the direction of the gravitational acceleration. We find that models of this type are disfavored relative to a simple dark matter halo model because the Milky Way data requires a substantial amplification of the radial acceleration with little amplification of the vertical acceleration.   We conclude that models which result in a MOND-like force struggle to simultaneously explain both the rotational velocity and vertical motion of nearby stars in the Milky Way.
\end{abstract}
\maketitle

\section{Introduction}
\label{sec:Intro}

In the late twentieth century, the discovery that rotation curves flatten at large radii~\cite{Rubin:1980zd,Bosma:1981zz} revolutionized our understanding of galactic dynamics. This singular observation is often considered the first definitive evidence for the missing mass problem. The standard resolution assumes that a (mostly) collision-less dark matter (DM) halo provides the required increase in acceleration. However, the lack of direct evidence for DM to date, together with numerous galactic-scale observations, provides an opening for other resolutions of this puzzle, such as long-distance, infrared  modifications to gravity as well as DM models that resemble such modifications on galactic scales.  In this paper, we present a general framework to test the standard DM solution against such alternatives.  One classic example of such an alternative to DM is MOdified Newtonian Dynamics (MOND)~\cite{Milgrom:1983pn,Milgrom:1983ca,Milgrom:1983zz}. As a first application of our approach, we demonstrate the tension of models that reduce to a MOND-like force with local Galactic observables, thereby challenging them precisely at the scale where they are designed to be successful.

Many aspects of the well-known observation of flat rotation curves have yet to be understood---a statement that remains true even when considering DM. Examples include the Baryonic Tully-Fisher relation~\cite{1977A&A....54..661T}, the Radial Acceleration Relation (RAR)~\cite{McGaugh:2016leg, 2000ApJ...533L..99M, 2001ApJ...563..694V} and additional correlations between baryons and observed accelerations on a galaxy-by-galaxy basis~\cite{Sancisi:2003xt,McGaugh:2014xfa,Lelli:2017vgz}. While such correlations are somewhat challenging to explain within a DM paradigm, some theories of modified gravity provide a natural explanation because they avoid the introduction of additional matter. While these models often suffer from lack of a consistent short-distance, ultraviolet counterpart~\cite{Bruneton:2007si} and struggle to explain cosmological observations~\cite{Milgrom:2014usa}, they are usually considered successful on galactic scales. MOND is probably the most well-known of these alternative explanations; it posits a modification to Newtonian gravity that becomes sizable only at extremely small accelerations.  Even if not a fundamental theory of nature, MONDian phenomenology is extremely successful on galactic scales. It is therefore imaginable that the true theory, be it DM, a modification to gravity or some hybrid, resembles MOND on these small scales. 
One example which has recently attracted attention is Superfluid DM~\cite{Berezhiani:2015bqa, Famaey:2017xou, Berezhiani:2017tth}. This model results in an emergent MOND-like force on galactic scales while remaining in the DM phase on larger scales. Other novel theories which resemble MOND include TeVeS~\cite{Bekenstein:2004ne}, Emergent Gravity~\cite{Verlinde:2010hp, Verlinde:2016toy}  and MOG~\cite{Moffat:2005si}. Throughout this paper we will refer to any theory which reduces to MOND-like phenomenology on galactic scales as a ``ML" model, even when the theory is not strictly a modification of gravity.  While this study is mostly agnostic to the details of any ML model, the specific case of Superfluid DM will be the focus of an upcoming paper.

In this work, we use local stellar dynamics to test DM against a general class of ML models. The Milky Way (MW) rotation curve is essentially constant at the Solar radius~\cite{2018arXiv181009466E}, where the naive, baryon-only prediction falls off quite rapidly.  As a result, local stellar measurements should be sensitive to the theory that governs the dynamics of the outer regions of the Galaxy~\cite{Famaey:2005fd,Stubbs:2005bv,Nipoti:2007sy, Bienayme:2009wb, Famaey:2011kh,Bienayme:2014kva,Margalit:2015zla}.  Although DM and ML models can both successfully explain the observed flatness of the MW rotation curve, they do so through different mechanisms that enhance acceleration. In a spherical DM halo, the increase in radial acceleration is due to additional mass surrounding the Galactic disk and is independent of the baryonic distribution. By contrast, in the ML framework considered in this study, the additional acceleration is determined purely by the baryonic distribution and is often in the form of an approximate scalar enhancement to the Newtonian acceleration, \textit{i.e.}, an enhancement that changes the magnitude, but not the direction, of the Newtonian acceleration. In particular, for a spherically symmetric DM distribution, the net vector enhancement of the acceleration always points towards the Galactic Center, while in a ML scenario, the vector enhancement is in the direction of the Newtonian acceleration that is expected from baryons only.

The ability of our analysis to differentiate between these models comes from the fact that we constrain the radial and vertical (perpendicular to the midplane) accelerations simultaneously. Importantly, measurements of the baryonic profile of the MW and of local velocity dispersions~\cite{Zhang:2012rsb} indicate that very little enhancement is required of the acceleration in the vertical direction.  On the other hand, measurements of the MW rotation curve~\cite{2018arXiv181009466E} require a much larger enhancement in the radial direction. Thus, a theory that treats both directions in an equivalent manner cannot easily explain all observations simultaneously, unless an anomalously large amount of baryonic mass is present at the center of the Galaxy. We find that  \emph{local observations point towards a model that enhances the radial acceleration without significantly affecting the vertical acceleration}. This provides a powerful handle to distinguish DM and any model that predicts a scalar-enhanced acceleration.

For the current study, we test ML forces using the local profile of the rotation curve, the local baryonic surface density, the vertical acceleration field within $\sim1$~kpc of the Galactic midplane, and the shape of the MW stellar and gas profiles. These are standard observables that are often used to constrain the MW potential and to estimate the local DM density---see Refs.~\cite{Bovy:2012tw, Garbari:2012ff, Smith:2011fs, Zhang:2012rsb,Bienayme:2014kva,Silverwood:2015hxa,2016MNRAS.458.3839X,Sivertsson:2017rkp} for the most recent studies, and~\Ref{Read:2014qva} for a review of the literature. We have performed a Bayesian likelihood analysis using these local MW observables to compare ML models and DM. In both cases, we model the MW baryonic density profile as a stellar disk, a gas disk, and a stellar bulge. We then fit these models to observations while marginalizing over uncertainties in the parameters. Tests of MOND using Galactic dynamics exist in the literature~\cite{Iocco:2015iia,Famaey:2005fd,Stubbs:2005bv,Nipoti:2007sy, Bienayme:2009wb, Famaey:2011kh, Bienayme:2014kva, Loebman:2014xha, Margalit:2015zla}, some of which have noted its effect on local vertical dynamics. However, a fully self-consistent study marginalizing over uncertainties in the baryonic distribution and directly comparing the results to a DM model has not been performed. Our analysis is also novel in that it is independent of any specific formulation of the ML model. Furthermore, our formalism has the potential to be extended to other scenarios.

We find that a theory with a ML force prefers a baryonic density profile that is in tension with known measurements of the stellar disk scale radius and the stellar bulge mass~\cite{Bland-Hawthorn:2016aaa,Licquia:2014rsa}. In particular, ML models require an anomalously small disk scale radius and/or an anomalously large bulge mass to reproduce local Galactic observables. On the other hand, DM is able to reproduce MW observables with parameters that are more consistent with the literature. Interestingly, the goodness of fit is slightly improved for a marginally prolate halo, \textit{i.e.}, one that introduces even less vertical acceleration than a spherical halo and is thus even further from the ML prediction.

Our results can be interpreted as model-independent in the sense that they need not rely on the details of the high and low-acceleration regimes in the Galaxy, but solely on the dynamics inferred from local measurements. When remaining completely agnostic to these details, the preference that we find for DM over ML models is positive, but not strong.\footnote{This study identifies observations where improvements in uncertainties can further strengthen the conclusions.} On the other hand we find \emph{a strong preference for DM over a ML force} for specific forms of the function which interpolates the high and low-acceleration regimes. One example of an interpolating function that is strongly constrained by our analysis is motivated by the recent study of rotation curves of $\sim 150$ galaxies in the SPARC database~\cite{Lelli:2016zqa}. Using this galaxy sample, Ref.~\cite{McGaugh:2016leg} finds the RAR, a universal relation between the observed radial acceleration and that expected due to baryons alone. The small scatter of the RAR can be interpreted as a manifestation of a ML force at play. Our result enables us to test the consistency of such an interpretation with Galactic observations.  We find that the interpolation function suggested by Ref.~\cite{McGaugh:2016leg} is inconsistent with the observed vertical acceleration of disk stars near the Solar position. This does not mean, however, that the RAR is in tension with our results more generally.

The paper is organized as follows. Sec.~\ref{sec: framework} describes the general framework proposed by this study, focusing on the differences between a DM model and a ML model. Sec.~\ref{sec: methodology} describes the Bayesian likelihood analysis that we perform. The modeling of the baryonic components are described in detail here, as well as the observational constraints that are used in the study.  Sec.~\ref{sec: results} presents the results of the analysis, explicitly demonstrating that the best-fit stellar parameters in the ML model are in tension with observations, and discussing the systematic uncertainties that affect these results in detail.  We conclude in Sec.~\ref{sec: conclusions}.   The Appendix supplements the discussion on non-linear effects in MOND and ML models and also includes further details on the likelihood analysis.

\section{Dark Matter vs. MOND-Like Models}
\label{sec: framework}
DM and ML models exhibit extremely different phenomenologies, even on galactic scales, due to their effects on the radial and vertical accelerations of tracer stars. In this section, we quantify this statement in further detail and point out specifically where the tension for ML models appears. Furthermore, we present details of our framework, which allows for strong distinguishing power between the different cases in a model-independent fashion. 

As discussed above, the constraining power arises from the fact that for a well-motivated baryonic profile, little additional vertical acceleration is required to explain observations, while a relatively large increase in the radial acceleration must be invoked.  The vertical acceleration is inferred from observations of the velocity dispersions of stars, while the radial acceleration is inferred from measurements of the local circular velocity of the MW.

DM is able to accommodate this requirement since an approximately spherical halo induces an acceleration pointing towards its center. Importantly, the net enhancement of the acceleration over the Newtonian contribution is just $\boldsymbol{a}_{\rm DM}$, the acceleration due to the DM component only, which is independent of the baryonic density profile. For a spherical DM profile, the local enhancement in cylindrical $(\hat{R},\hat{z})$ coordinates is $\boldsymbol{a}_{\rm DM} \approx -G M(R_0)/R_0^2 \times (1,z/R_0)$, where $M(R_
0)$ is the enclosed DM mass, $z$ is the height above the midplane and $R_0$ is the radius of the Solar position. Thus, most of the local enhancement occurs in the radial direction with the vertical enhancement suppressed by $\sim z/R_0$.

The situation is quite different in a ML scenario, where the galactic dynamics are driven solely by the baryonic distribution. For many such models, the response to matter is highly non-linear, making a prediction of the dynamics at a given point within the galaxy non-trivial to calculate. However, as we demonstrate in this work, one need only characterize some general properties of the gravitational response to matter to provide discriminating power between ML models and DM. The particular characteristics that are important to classify are:  (1)~the parametric functional relationship between the local dynamical acceleration, $\boldsymbol{a}$, and the baryonic matter distribution, $\rho_{\rm B}$; (2)~the tensor structure of this function; and (3)~the degree to which this function varies in the region of interest within the MW.

For example, in many formulations of ML models, the observed acceleration, $\boldsymbol{a}$, depends only on $\rho_{\rm B}$ via the Newtonian acceleration, $\boldsymbol{a}_{\rm N}$. The asymptotic behavior is designed to reproduce the observed flatness of rotation curves in galaxies and is determined by
\begin{equation}
a = 
\begin{cases} 
      a_{\rm N} & a \gg a_0 \\
      \sqrt{a_0 a_{\rm N}} & a \ll a_0 \, ,
\end{cases}
\label{eq:a_MOND}
\end{equation}
where $a_0$ is a constant acceleration scale that sets the deviation from Newtonian gravity.
In general, the solution for $\boldsymbol{a}$ is model-dependent and difficult to obtain.
However, under certain conditions (see \App{sec:non_linear_effects_in_mond}), the dynamics reduces to the following form:
\beq
\boldsymbol{a} = \nu\left(\frac{a_{\rm N}}{a_0}\right) \boldsymbol{a}_{\rm N} \, .
\label{eq:nu_aobs}
\eeq
The function $\nu(a_{\rm N}/a_0)$ is known as the interpolation function and satisfies the asymptotic conditions of \Eq{eq:a_MOND}. Thus, for the case of a ML model, Eq.~(\ref{eq:nu_aobs}) is the parametric functional relationship between the dynamical acceleration and the baryonic distribution and it manifests as a scalar enhancement to the Newtonian acceleration.  Importantly, in vanilla MOND, the specific form of this function is arbitrary, while in other models, \textit{e.g.} Superfluid DM, it is set by the parameters of the model and/or the baryonic profile~\cite{Berezhiani:2017tth}.

Finally, if one is only interested in fitting a ML model to Galactic observables in a localized region near the Sun, then the Newtonian acceleration, $a_{\rm N}$, does not vary too much from some particular reference value, $a_{\rm N, ref}$. Therefore, one can expand $\nu(a_{\rm N}/a_0)$ in a Taylor series, giving
\begin{eqnarray}
  \nu\left(\frac{a_{\rm N}}{a_0}\right) &=& \nu\left(\frac{a_{\rm N,ref}}{a_0}\right) + \nu'\left(\frac{a_{\rm N,ref}}{a_0}\right)\frac{a_{\rm N} - a_{\rm N,ref}}{a_0} \nonumber \\ 
  & \equiv& \nu_0 + \nu_1 \cdot a_{\rm N} 
\label{eq:invmu}
\end{eqnarray}
to first order, where we have parametrized in terms of the constants $\nu_0\equiv\nu(a_{\rm N,ref}/a_0) - \nu'(a_{\rm N,ref}/a_0)\cdot a_{\rm N,ref}/a_0$ and  $\nu_1\equiv \nu'(a_{\rm N,ref}/a_0)/a_0$.\footnote{Note that in this parametrization, one can only constrain the value of $\nu_1$, \emph{i.e.}, $\nu'(a_{\rm N,ref}/a_0)/a_0$, and not the scale $a_0$ itself.}
In doing so, the arbitrary function in \Eq{eq:nu_aobs} is reduced to two constants, which can be treated as free parameters of the model. This expansion allows one to test any ML model and present results in a model-independent manner.

\begin{figure}[t] 
\centering
\includegraphics[width=3.5in]{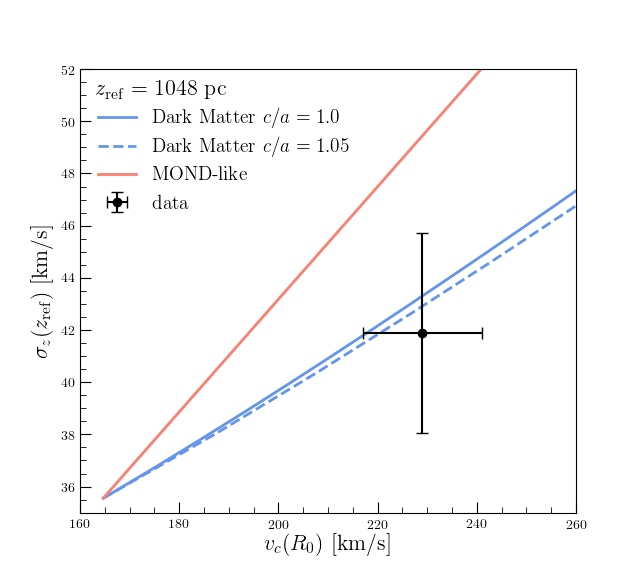} 
\caption{Illustrative plot presenting the potential ability of dark matter and a MOND-like model to predict the measured values of the circular velocity at the Solar position, $v_c(R_0)$, and the vertical velocity dispersion at some reference height above the midplane,  $\sigma_z(z_{\rm ref})$. For this figure, the baryonic profile has been fixed such that it is consistent with measurements of the stellar disk and bulge, as well as the gas disk. The \textit{black point} marks the measurements taken from Refs.~\cite{2018arXiv181009466E} and~\cite{Zhang:2012rsb} (see text for details). The \textit{blue solid curve} is the prediction for dark matter with a spherical NFW profile and the \textit{dashed blue curve} is the prediction for dark matter with a slightly prolate profile. The \textit{red solid curve} is the prediction for a MOND-like force using Eqs.~(\ref{eq:nu_aobs}) and~(\ref{eq:invmu}). The MOND-like model requires equal enhancements of the radial and vertical accelerations and cannot simultaneously fit both.  On the other hand, both dark matter scenarios are able to accommodate the measurements. The prolate halo does slightly better because it increases the enhancement of the radial acceleration relative to the vertical acceleration, a feature that is the exact opposite of the MOND-like behavior.
}
\label{fig:cartoon}
\end{figure}

Given the phenomenology of each of these types of models, one can now ask which is better suited to simultaneously fit observations of the local radial and vertical accelerations. Measurements of these values are often inferred indirectly from the local circular velocity, $v_c(R_0)$, which correlates with the radial acceleration through $a_R=v_c^2/R_0$, and from the vertical velocity dispersion, $\sigma_z(z)$, which correlates with the vertical acceleration through the Jeans equation~\cite{2008gady.book.....B}. \Fig{fig:cartoon} presents the different predictions of these values for DM versus a ML model for $v_c(R_0)$ and for $\sigma_z(z_{\rm ref})$ at a reference value of $z_{\rm ref}=1048$ pc.\footnote{The measurement of $\sigma_z(z_{\rm ref})$ used in this plot corresponds to the measurement of the metal-poor tracer population at $z_{\rm ref}=1048$~pc taken from Ref.~\cite{Zhang:2012rsb}, however the qualitative results shown in the figure are independent of these choices.}  For the purposes of illustration, we take a fixed baryonic profile that is consistent with measurements of the stellar disk and bulge as well as the gas disk. The black point indicates the measured values from Refs.~\cite{2018arXiv181009466E} and~\cite{Zhang:2012rsb}.

The solid blue curve in the figure indicates the prediction for a DM model consisting of a spherical Navarro-Frenk-White~(NFW) halo~\cite{Navarro:1995iw}, varying over its normalization. Clearly, given this baryonic profile, the DM model easily accommodates the measured $v_c(R_0)$ and $\sigma_z(z_{\rm ref})$ simultaneously. The dashed blue curve indicates the same correlation for a halo that is slightly prolate.  We take the halo model of Ref.~\cite{Bovy:2016chl} with their best-fit axis ratio of $c/a=1.05$. Compared to the spherical case, this model enhances $v_c(R_0)$ for a given $\sigma_z(z_{\rm ref})$ and is therefore slightly closer to the central values of the measurements.

The red solid curve in the figure is the correlation of these values for a ML model using \Eq{eq:nu_aobs} and the approximation for $\nu(a_{\rm N}/a_0)$ in \Eq{eq:invmu}, varying over $\nu_0+\nu_1 a_{\rm N}$. This curve shows that such models are unable to fit the measured values simultaneously. Specifically, in order to fit to $\sigma_z(z_{\rm ref})$, they under-predict $v_c(R_0)$. One way around this is to invoke a different baryonic potential with much more mass at the center of the Galaxy. However, as is shown in the rest of this study, this is in conflict with measurements of the MW bulge mass and stellar disk scale radius.

The example presented in this section is meant for simple illustration of the primary tension underlying the ML and DM scenarios.  To draw a more robust conclusion, one should marginalize over the baryonic uncertainties, which we describe in full in the next few sections.

Lastly, we comment that for ML models, the general relation between observed and Newtonian acceleration is slightly more complex than \Eq{eq:nu_aobs}. Additionally, the non-linear structure of these theories can potentially cause further complications regarding the validity of the assumed smoothness of the baryonic profile. However, as is detailed in \App{sec:non_linear_effects_in_mond}, under certain circumstances, deviations from these assumptions are expected to be negligible. We find this to be the case for local measurements and thus for all results in this study. We stress that our analysis can be extended to more general frameworks; however, in this paper, we will restrict our focus to ML scenarios that satisfy \Eq{eq:nu_aobs}.

\section{Methodology}  
\label{sec: methodology}
This section describes the Bayesian likelihood analysis that we perform to test the consistency of the DM and ML models with local MW observables.
For a given model $\mathcal{M}=\rm{DM},\rm{ML}$ with parameters $\boldsymbol{\theta}_{\mathcal{M}}$, we can predict the values of different MW observables $\mathbf{X}(\boldsymbol{\theta}_{\mathcal{M}}) = (X_{1}(\boldsymbol{\theta}_{\mathcal{M}}), ..., X_{N}(\boldsymbol{\theta}_{\mathcal{M}}))$.  The predicted values are then compared to the measured values $\mathbf{X}_{\rm{obs}}=(X_{1,\rm{obs}}, ..., X_{N,\rm{obs}})$ with uncertainties $\delta\mathbf{X}_{\rm{obs}}=(\delta X_{1,\rm{obs}}, ..., \delta X_{N,\rm{obs}})$.  We define the likelihood function to be  
\beq
\mathcal{L}(\boldsymbol{\theta}_{\mathcal{M}}) \propto \exp\left[-\frac{1}{2}\sum_{j=1}^{N} \left(\frac{X_{j,\rm{obs}} - X_{j}(\boldsymbol{\theta}_{\mathcal{M}})}{\delta X_{j,\rm{obs}}}\right)^2\right] \, 
\label{eq:likelihood}
\eeq
and seek to recover the posterior distributions of the model parameters using Bayesian inference.\footnote{Note that in this equation we have assumed that the measurements are uncorrelated. Since most of the observations are taken from entirely different data sets, the correlations are expected to be small. An analysis which accounts for the full covariance matrix is beyond the scope of this study.}

In \Sec{sec:models}, we define the models, in \Sec{sec: constraints} we describe the observables $\mathbf{X}_{\rm obs}$ that are used as constraints, and in \Sec{sec:mcmc} we discuss the analysis procedure in detail.

\subsection{Model Definition}
\label{sec:models}
In this section, we detail the parametrization of the DM and ML models.
In both cases, the predicted acceleration depends sensitively on the MW baryonic mass density, $\rho_{\rm B}$, which we take to consist of a stellar bulge and a disk of stellar ($*$) and gaseous ($g$) components:
\beq
\rho_{\rm B} = \rho_{*,\rm{bulge}} + \rho_{*,\rm{disk}} + \rho_{g,\rm{disk}} \, .
\eeq
For both the stellar and gaseous disks, we use a double-exponential density profile given by
\beq
\rho_{j,{\rm disk}}(R, z) = \tilde{\rho}_{j} \,  \exp\left(-R/h_{j,R} - |z|/h_{j,z} \right),
\label{eq:expDisk}
\eeq
where $h_{j,R}$ ($h_{j,z}$) is the scale length (height) of the disk, $\tilde{\rho}_{j}$ is the normalization, and the index $j=*,g$ corresponds to either the stellar or gas component. The variables $R$ and $z$ are Galactocentric cylindrical coordinates.  

For computational simplicity, we do not separately model the thin and thick stellar disks, but rather treat these contributions together as a single exponential density profile with an effective scale height and radius.  As the local density ratio of the thick to thin disk ranges from $\sim 1$--10\%~\cite{Gilmore:1983bv, Siegel:2002vr, Juric:2005zr}, the properties of the effective disk are primarily set by the thin component.  We note that the effective disk can also be considered to be the weighted contribution of individual mono-abundance populations~\cite{Bovy:2011ux,Bovy:2013raa}, which yields roughly the same effective scale length at the Solar position compared to what we estimate from the thin/thick disk contributions.

In principle, there are 6 free parameters for the disk model; however, we restrict half of these values. For our benchmark analysis, we take  $h_{g, z} = 130$~pc and $h_{g, R} = 2 \, h_{*, R}$, following \Refs{2008gady.book.....B,Bovy:2013raa}, and also fix the stellar scale height to $h_{*,z} = 300$~pc, consistent with measurements from stellar counts~\cite{Bland-Hawthorn:2016aaa}.

For the stellar bulge, we use a Hernquist density profile~\cite{Hernquist:1990be},
\beq
\rho_{*,{\rm bulge}}(r) = \frac{M_{*,\rm bulge}}{2\pi}\frac{r_{*,\rm bulge}}{r}\frac{1}{(r+r_{*,\rm bulge})^3},
\label{eq:Hernquist}
\eeq
where $r_{*,\rm bulge}$ is the bulge scale radius, and $M_{*,\rm bulge}$ is a normalization constant that sets the total mass of the bulge.  Here, $r$ is the Galactocentric spherical radial coordinate. The scale radius is fixed to $r_{*,\rm bulge} = 600$~pc, while the normalization is allowed to vary.  While the MW bulge has more structure than captured by \Eq{eq:Hernquist} (see \Ref{Bland-Hawthorn:2016aaa} for a review), our analysis is not sensitive to these details as the observables of interest are measured beyond $R \gtrsim 5$~kpc.

Finally, we must specify the dynamics of the DM and ML models. For the DM scenario, the dynamical acceleration is given by the solution to Poisson's equation,
\beq
\boldsymbol{\nabla}\cdot\mathbf{a} = -4\pi G (\rho_{\rm B} +\rho_{\rm DM}),
\label{eq: DMa}
\eeq
where $\rho_{\rm DM}$ is the DM mass density.
In particular, we assume that the DM density profile follows a generalized NFW distribution~\cite{Navarro:1995iw},
\beq
\rho_{\rm DM}(r) = \frac{\tilde{\rho}_{\rm DM}}{\left(r/r_s\right)^{\alpha} \left(1+r/r_s\right)^{3-\alpha}} , 
\label{eq:rho_DM}
\eeq
where $r_s$ is the scale radius, and $\alpha$ is the inner slope. The parameters $\tilde{\rho}_\mathrm{DM}$ and $\alpha$ are both free in our analysis procedure, while we fix $r_s=19$~kpc.

On the other hand, for the ML scenario in our benchmark analysis, the observed dynamical acceleration is given by
\beq
\mathbf{a} = \left(\nu_0 + \nu_1 a_{\rm N}\right)\mathbf{a}_{\rm N},
\label{eq: aG}
\eeq
where $\nu_0$ and $\nu_1$, defined in \Eq{eq:invmu}, parametrize the interpolation function and $\mathbf{a}_{\rm N}$ is the Newtonian acceleration, which satisfies
\beq
\boldsymbol{\nabla}\cdot\mathbf{a}_{\rm N} = -4\pi G \rho_{\rm B}.
\eeq
We allow $\nu_0$ and $\nu_1$ to be free parameters in the analysis procedure; however, we restrict the model to ensure that the observed acceleration is always an enhancement of Newtonian gravity by forcing ${\nu_0+\nu_1 a_{\rm N}>1}$ everywhere within the spatial region of interest.

In addition to this benchmark analysis, we also consider a number of specific interpolation functions, as well as the use of the interpolation function of Eq.~\eqref{eq: aG} with the requirement that ${\nu_0+\nu_1 a_{\rm N}>1.3}$. Details of all functions are presented in Table~\ref{tab:nu_funcs}. For the interpolation functions which are not Taylor expansions, there is a single free parameter, $a_0$, for which we set priors as detailed in the table. The results of these analyses are far less general than the use of Eq.~\eqref{eq: aG}, but are more in line with specific formulations of MOND and ML models, including Superfluid DM. Generically, they force $\mathbf{a} > \mathbf{a}_{\rm N}$ at the Solar position, whereas with Eq.~\eqref{eq: aG} and the requirement ${\nu_0+\nu_1 a_{\rm N}>1}$, the model also allows $\mathbf{a} \approx \mathbf{a}_{\rm N}$ at the Solar position.

\begin{table*}
\begin{center}
\renewcommand{\arraystretch}{3}
\footnotesize
\begin{tabular}{C{3cm}|C{6cm}C{5cm}C{3cm}}
  \Xhline{3\arrayrulewidth}
Naming Convention & Functional Form & Prior for Scan & $\Delta$BIC \\
\hline
Taylor Expansion & $\nu(a_{\rm N}) = \nu_0 + \nu_1 a_{\rm N}$ & $\nu(a_{\rm N}) > 1 \text{ or } 1.3$ & 4.1 or 7.5\\
RAR~\cite{McGaugh:2016leg} & $\nu(a_{\rm N}) = \left( 1 - e^{-\sqrt{a_{\rm N}/a_0}} \right)^{-1}$ & $a_0=\mathrm{LOGNORMAL}\left(1.20, 0.24^2\right)$ & 10.4 \\
Simple~\cite{Hees:2015bna,Famaey:2011kh} & $\nu(a_{\rm N}) = \frac{1}{2} \left( 1 + \sqrt{1 + \frac{4}{a_{\rm N}/a_0}} \right)$ &  $a_0=\mathrm{LOGNORMAL}\left(1.2, 0.4^2\right)$ & 9.6 \\
Standard~\cite{Hees:2015bna,Famaey:2011kh}& $\nu(a_{\rm N}) = \sqrt{\frac{1}{2} \left( 1 + \sqrt{1 + \left(\frac{2}{a_{\rm N}/a_0} \right)^2} \right)}$ &  $a_0 = \mathrm{LOGNORMAL}\left(1.2, 0.4^2\right)$ & 4.8 \\
\Xhline{3\arrayrulewidth}
\end{tabular}
\end{center}
\caption{Details of interpolation functions used in this study. Naming conventions, functional forms, priors and the values of $\Delta$BIC are given. The prior for the RAR function is taken directly from the results of fitting the function to the RAR in \Ref{McGaugh:2016leg}, while the priors for the Simple and Standard functions are taken to have the same central value for $a_0$, 
but with a somewhat larger variance to account for the fact that this value of $a_0$ is obtained by fitting with a different function. Note that here the notation $\mathrm{LOGNORMAL}(m,v)$ denotes a lognormal distribution with mean $m$ and variance $v$. Values of $\Delta$BIC are given for a bulge mass $M_{*, {\rm bulge,obs}} = 1.50\pm0.38\times10^{10} M_{\odot}$. }
\label{tab:nu_funcs}
\end{table*}

\subsection{Constraints from Local Observations}
\label{sec: constraints}

Next, we discuss the values of the observables that are used in the likelihood analysis.  The solar radius, $R_0$, feeds into the prediction for each of these observables.\footnote{In this work, we neglect the vertical offset of the Sun above the Galactic plane, \textit{i.e.}, we take $\zsun=0$.}  This parameter is critical because any Galactic measurements based on angular size or velocity perceived from Earth depend on $R_0$.  We take $R_0 = 8.122$~kpc as the fiducial value, consistent with the observation of the orbit of the star S2 around the massive black hole candidate Sgr A*~\cite{Abuter:2018drb}. 

The first constraint arises from the local stellar and gas surface densities, as determined from photometric observations.  By definition, the surface density depends on the  distribution of the stellar or gas component as
\beq
\Sigma_j^{z_{\rm max}} = 2 \int_0^{z_{\rm max}} \rho_j\left(R_0,z'\right) \ dz'\ .
\label{eq: Sigma11}
\eeq
We adopt measurements of ${\Sigma_{g, \text{obs}}^{1.1} = 12.6 \pm 1.6\ \mathrm{M}_\odot\ \mathrm{pc}^{-2}}$ and ${\Sigma_{*, \text{obs}}^{1.1} = 31.2 \pm 1.6\ \mathrm{M}_\odot\ \mathrm{pc}^{-2}}$ at ${z_{\rm max}=1.1\ \mathrm{kpc}}$ as fiducial values~\cite{McKee:2015hwa}.
For self-consistency, we only use surface densities obtained from direct photometric observations (as opposed to dynamical studies), which are equally valid for the DM and ML scenarios.

The second constraint comes from the value and slope of the rotation curve at the Solar radius.  The circular velocity and its derivative with respect to $R$ are obtained from the predicted acceleration using
\beq
v_c(R) = \left.\sqrt{R\cdot a(R)}\right|_{z=0} \, .
\label{eq:vcirc}
\eeq
We consider only the local circular velocity and slope of the rotation curve at $R=R_0$, taking $v_{c, \text{obs}} = 229 \pm 12 ~\rm km~sec^{-1}$ and $\left(dv_c/dR\right)_{\rm obs} = -1.7 \pm 0.47~\rm km~sec^{-1}~kpc^{-1}$ as fiducial values~\cite{2018arXiv181009466E}. Note that we have added in quadrature the statistical and systematic uncertainties quoted in~\Ref{2018arXiv181009466E} and, in particular, that the $\sim 5\%$ systematic uncertainties in $v_c$ dominate. 

\begingroup
\squeezetable
\begin{table*}
\centering
\begin{tabular}{C{1.4cm}|C{1.4cm}C{1.2cm}C{1.2cm}C{1.2cm}C{1.2cm}C{1.2cm}C{1.2cm}C{1.2cm}C{1.2cm}C{1.6cm}C{2cm}}
  \Xhline{3\arrayrulewidth}
\renewcommand{\arraystretch}{1}
  & $\tilde{n}_i$ & $h_i$ & $\tilde{\rho}_*$ & $h_{*,R}$ & $M_{*,\rm bulge}$ & $\tilde{\rho}_g$ & $\tilde{\rho}_{\rm DM}$ & $\alpha$ & $\nu_0$ & $-\nu_1$ & $a_0$ \\
\hline
Unit & $10^{-3}\ \mathrm{pc}^{-3}$ & kpc & $\mathrm{M}_{\odot}\ \mathrm{pc}^{-3}$ & kpc & $10^{10}\ \mathrm{M}_{\odot}$ & $\mathrm{M}_{\odot}\ \mathrm{pc}^{-3}$ & $\mathrm{M}_{\odot}\ \mathrm{pc}^{-3}$ & -- & -- & $10^{10}\ \mathrm{s}^2\ \mathrm{m}^{-1}$ & $10^{-10}\ \mathrm{m}\ \mathrm{s}^{-2}$ \\

Prior & [$10^{-2}$, 1] & [$10^{-3}$, 8] & [0, $10^2$] & [$10^{-3}$, 8] & [0, $10^{2}$] & [0, $10^2$] & [0, $10^2$] & [0, $10^2$] & [0, $10^2$] & [0, $10^2$] & LN$(1.20,0.24^4)$\\
Posterior (ML) & $34.7^{+5.7}_{-4.8}$ $4.43^{+0.48}_{-0.41}$ $2.07^{+0.13}_{-0.12}$ & $0.26^{+0.01}_{-0.01}$ $0.46^{+0.03}_{-0.03}$ $0.87^{+0.06}_{-0.05}$ & $1.37^{+1.87}_{-0.95}$ & $2.41^{+1.26}_{-0.48}$ & $4.29^{+2.36}_{-2.37}$ & $0.25^{+0.14}_{-0.11}$ & -- & -- & $1.44^{+0.13}_{-0.08}$ & $0.21^{+0.05}_{-0.03}$ & --\\
Posterior (DM) & $33.6^{+5.4}_{-4.6}$ $4.35^{+0.44}_{-0.39}$ $2.06^{+0.13}_{-0.12}$ & $0.26^{+0.01}_{-0.01}$ $0.46^{+0.03}_{-0.03}$ $0.87^{+0.06}_{-0.05}$ & $1.21^{+1.17}_{-0.81}$ & $2.54^{+1.29}_{-0.45}$ & $2.87^{+2.12}_{-1.82}$ & $0.23^{+0.10}_{-0.10}$ & $2.16^{+1.12}_{-1.06}$ & $0.40^{+0.54}_{-0.30}$ & -- & -- & -- \\
Posterior (ML RAR) & $39.8^{+6.5}_{-5.4}$ $4.8^{+0.48}_{-0.43}$ $2.14^{+0.13}_{-0.12}$ & $0.25^{+0.01}_{-0.01}$ $0.44^{+0.02}_{-0.02}$ $0.82^{+0.05}_{-0.05}$ & $0.56^{+0.52}_{-0.31}$ & $3.12^{+1.33}_{-0.62}$ & $4.10^{+1.05}_{-1.46}$ & $0.16^{+0.07}_{-0.05}$ & -- & -- & -- & -- & $0.91^{+0.14}_{-0.13}$\\
\Xhline{3\arrayrulewidth}
\end{tabular}

\caption{\label{tab:priors} Free parameters used in the benchmark Bayesian likelihood analysis along with the associated prior range for each and the 16-50-84$^\text{th}$ percentiles of the marginalized posterior. From left to right: normalization and scale height of $i^\text{th}$ tracer population, normalization and scale length of stellar disk, bulge mass, normalization of gas disk, dark matter density normalization and inner slope, and interpolation function parameters for the ML models.  The notation for the prior functions is: $[a,b]$ denotes a flat prior between $a$ and $b$; $\mathrm{LN}(m,v)$ denotes a lognormal distribution with mean $m$ and variance $v$.}

\end{table*}
\endgroup

The final constraint comes from the observed number density and vertical velocity dispersions of three mono-abundance stellar populations at $R=R_0$, provided in~\Ref{Zhang:2012rsb}.  These results were obtained using 9000 K-dwarfs in the SEGUE sub-survey of the Sloan Digital Sky Survey (SDSS)~\cite{York:2000gk, Abazajian:2008wr, Aihara:2011sj}.  The sample was categorized based on its iron fraction, [Fe/H], and $\alpha$-abundance, [$\alpha$/Fe], and divided into three tracer populations consisting of metal-rich, metal-intermediate, and metal-poor stars.  For each population, indexed by $i$, number densities, $n_{i, {\rm obs}}(z_k)$, and vertical velocity dispersions, $\sigma_{z,i, {\rm obs}}(z_k)$, were obtained for several values of $z_k$ between $300~\mathrm{pc}$ and $1200\ \mathrm{pc}$.

Following~\Ref{Zhang:2012rsb}, we model the number densities, $n_i(z)$, for the $i^\text{th}$ tracer population as
\beq
n_i(z) = \tilde{n}_i \, \text{exp}(-\left|z\right|/h_i) \, , 
\label{eq: num_dens}
\eeq
where $\tilde{n}_i$ and $h_i$ are additional parameters varied in the fit.  The tracer stars are assumed to be in steady state and to be well-described by the Jeans equations~\cite{2008gady.book.....B}.  In this case, the vertical velocity dispersion is
\bea
\sigma_{i,z}(z)^2  &=& \frac{-1}{n_i(z)} \int_z^\infty n_i(z^\prime) \, a_z(z^\prime) \, dz^\prime \, ,
\label{eq:sigz} 
\eea
where $a_z$ is the predicted vertical acceleration. Although the integral in \Eq{eq:sigz} extends to infinity, the integrand falls off exponentially in $z$ due to \Eq{eq: num_dens}, and we are not sensitive to dynamics above $z \gtrsim h_i\sim 1\ \mathrm{kpc}$. Additionally, \Eq{eq:sigz} ignores the contribution from the tilt term, which depends on the velocity dispersion in the $R - z$ plane.  The contribution of this term is negligible, as estimated in~\Ref{Zhang:2012rsb} for the specific K-dwarf sample studied there, and can thus be safely neglected. 

\subsection{MCMC Analysis}
\label{sec:mcmc}

To summarize, the complete set of model parameters is formed by the union of the set of SEGUE tracer population parameters, $\{\tilde{n}_i,\,h_i\,|\,i~=~1,2,3\}$, the baryon potential parameters, $\{\tilde{\rho}_*,\ h_{*,R},\ h_{*,z},\ M_{*,\rm bulge},\ r_{*,\rm bulge},\ \tilde{\rho}_g,\  h_{g,R},\ h_{g,z}\}$, and $\{\nu_0, \nu_1\}$ (or $a_0$) for the ML model or $\{\tilde{\rho}_\text{DM}, \alpha, r_s\}$ for the DM case.  We set $r_s$, $r_{*,\rm bulge}$, $h_{*,z}$, and $h_{g,z}$ to constant values, and fix $h_{g,R}=2h_{*,R}$. It would be preferable to marginalize over all parameters in the scans, however this is computationally prohibitive. The effects of varying these fixed values are summarized in detail in Sec.~\ref{sec:uncertainties}.  
The free parameters in \Eq{eq:likelihood} are therefore
\bea
\boldsymbol{\theta}_{\rm ML} &=&  \left(\tilde{n}_i,\ h_i,\ \tilde{\rho}_*,\ h_{*,R},\ M_{*,\rm bulge},\ \tilde{\rho}_g,\ \nu_0,\ \nu_1 \right) \\
\boldsymbol{\theta}_{\rm DM} &=& \left(\tilde{n}_i,\ h_i,\ \tilde{\rho}_*,\ h_{*,R},\ M_{*,\rm bulge},\ \tilde{\rho}_g,\ \tilde{\rho}_\text{DM},\ \alpha\right), \nonumber
\eea
(for scans with specific interpolation functions $\{\nu_0,\ \nu_1\} \to a_0$) and the local observables are
\bea
\textbf{X}_{\rm obs} = && \left(n_{i,\rm obs}(z_k),\ \sigma_{z,i,\rm obs}(z_k),\ \Sigma_{*,\rm obs}^{1.1},\ \Sigma_{g,\rm obs}^{1.1}, \right. \nonumber \\
&& \quad \left.\ v_{c,\rm obs},\ \left(dv_c/dR\right)_{\rm obs}\right)  \, .
\eea
For a given model, $\mathcal{M}$, the predicted values of these observables, $\textbf{X}(\boldsymbol{\theta}_{\mathcal{M}})$, are obtained from Eqs.~\eqref{eq: Sigma11}-\eqref{eq:sigz}.

We use the Affine Invariant Markov Chain Monte Carlo (MCMC) implementation \texttt{emcee}~\cite{2013PASP..125..306F} to recover the posterior distributions for $\boldsymbol{\theta}_{\rm ML}$ and $\boldsymbol{\theta}_{\rm DM}$, as appropriate, using the likelihood defined in \Eq{eq:likelihood}.  For our benchmark analysis, we set flat linear priors on each parameter as summarized in Table~\ref{tab:priors} and additionally require that each walker step satisfy (i)~$\nu_0 + \nu_1 a_{\rm N} > 1$ and (ii)~that the baryonic rotation curve peak below $R < 5$~kpc.\footnote{This is motivated since the MW is known to have a maximal disk~\cite{Sackett:1997wf}, with the rotation curve peak occurring in a region that is baryon-dominated.}   For the additional analyses, we set either flat or log-normal priors as detailed in Table~\ref{tab:nu_funcs}.  The details of the MCMC analysis, including posterior distributions, two-parameter correlations, and convergence, are presented in App.~\ref{sec:mcmc_convergence}.

\begin{figure*}[htbp] 
\centering
\includegraphics[width=.34\textwidth]{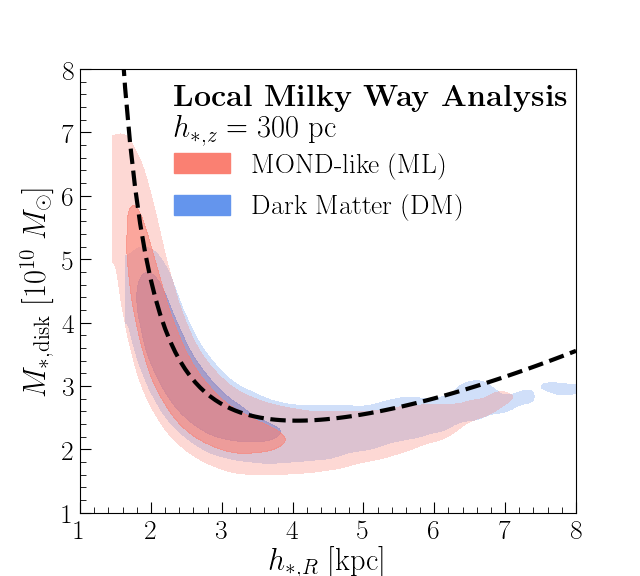}\includegraphics[width=.34\textwidth]{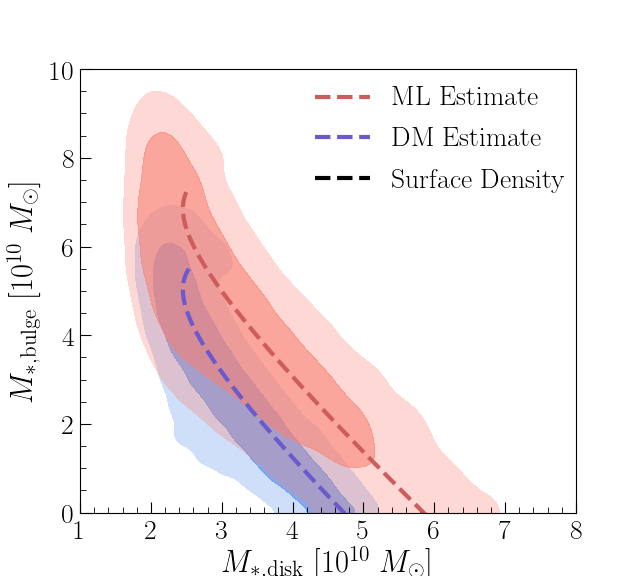}\includegraphics[width=.34\textwidth]{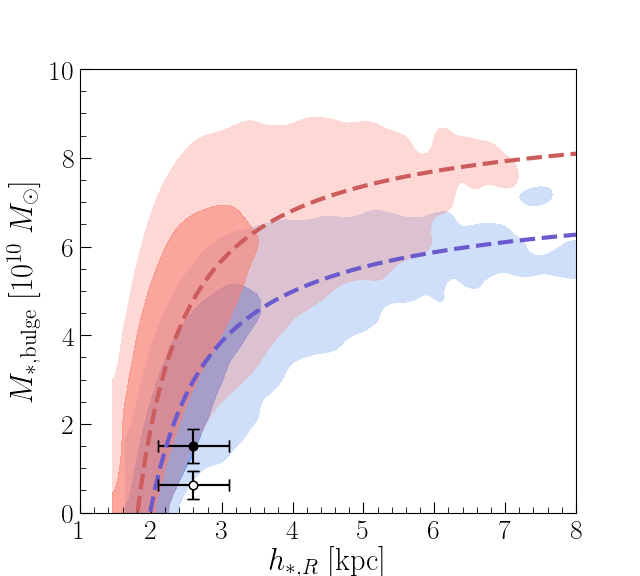}
\caption{Marginalized two-parameter correlations for the disk and bulge parameters $h_{*,R}$, $M_{*,{\rm disk}}$, and $M_{*,{\rm bulge}}$.  Shaded regions correspond to 68\% and 95\% containment regions for the MOND-like model (red) and dark matter (blue) models.  {\bf Left panel}: Correlation between stellar disk mass, $M_{*,{\rm disk}}$, and scale radius, $h_{*,R}$.  The \textit{black dashed curve} shows the range of parameters consistent with the measured stellar surface density $\Sigma_{*, \text{obs}}^{1.1} = 31.2$~$M_\odot$~pc$^{-2}$~\cite{McKee:2015hwa}, as in \Eq{eq:Mdisk_HR_correlation}. {\bf Central panel}: Correlation between stellar bulge mass, $M_{*,{\rm bulge}}$, and stellar disk mass, $M_{*,{\rm disk}}$. The {\it red dashed curve} corresponds to the correlation that reproduces the measured local circular velocity assuming Eq.~(\ref{eq:Mdisk_HR_correlation}) and a constant value of $\nu(a_{\rm N}/a_0) = 1.3$ for a MOND-like model. The {\it blue dashed curve}  is the analogous estimate for the dark matter model using $\tilde{\rho}_{\rm DM}=2.8\ M_{\odot}\ \mathrm{pc}^{-3}$ and $\alpha=0.5$. {\bf Right panel}: Correlation between $h_{*, R}$ and $M_{*,{\rm bulge}}$. The {\it red and blue dashed curves} are analogous to those in the central panel.  The \textit{open/filled black points} correspond to two separate bulge measurements:  $M_{*, {\rm bulge,obs}} = 1.50 \pm 0.38$~$M_\odot$ from microlensing observations~\cite{Novatin:2007dd} (filled black circle) and  $M_{*, {\rm bulge,obs}} = 0.62 \pm 0.31$~$M_\odot$ (open black circle) from photometric observations~\cite{LopezCorredoira:2006na}. We take $h_{*, R, {\rm obs}} = 2.6\pm0.5$~kpc~\cite{Bland-Hawthorn:2016aaa} for both. Compared to the MOND-like model, the dark matter region has greater overlap with the favored parameter space.}
\label{fig:Fits_1}
\end{figure*}

\section{Results}  
\label{sec: results}
\subsection{The Baryonic Density Profile}
\label{sec:baryons}

This section presents the results of the Bayesian parameter inference described in \Sec{sec: methodology}.  The median values and uncertainties of all the free parameters of our baseline scans (as well as the RAR function scan for comparison) are presented in Table~\ref{tab:priors}. We start by exploring the preferred baryonic mass distributions for both the DM and ML scenarios. The results are summarized in Fig.~\ref{fig:Fits_1}, which shows the correlations between the stellar disk scale radius $h_{*,R}$, stellar disk mass $M_{*,{\rm disk}}$, and bulge mass $M_{*,{\rm bulge}}$. Note that, as will be discussed below, in the final results presented at the end of this section, the scale radius and bulge mass are set as constraints in the analysis. However, the correlations shown in the figure best illustrate the result of this study: the DM model is more consistent with observations than the ML model. The 68\% and 95\% containment regions are indicated for both the ML~(red) and DM~(blue) models.

The left panel of \Fig{fig:Fits_1} shows the correlation of $h_{*,R}$ with $M_{*,{\rm disk}}$, for a specific value of the stellar scale height, $h_{*,z} = 300$~pc. The correlation between these two parameters is mostly driven by the constraint on the local stellar surface density. Using \Eqs{eq:expDisk}{eq: Sigma11}, it can easily be shown that
\beq
M_{*,{\rm disk}} = \frac{2\pi \, h_{*,R}^2 \, \Sigma_{*,{\rm obs}}^{z_{\rm max}} \, \exp(R_0/h_{*,R})}{1 - \exp(-z_{\rm max}/h_{*,z})},
\label{eq:Mdisk_HR_correlation}
\eeq
where $z_{\rm max}=1.1$~kpc for the constraints used in this study.  The resulting curve is indicated by the black dashed line; both the ML and DM best-fit regions roughly follow this trend. 

The central panel of Fig.~\ref{fig:Fits_1} illustrates the correlation of the stellar bulge mass, $M_{*,{\rm bulge}}$, with the stellar disk mass, $M_{*,{\rm disk}}$. This correlation is driven primarily by the value of the local circular velocity, which is roughly determined by the total enclosed baryonic mass.
Therefore, in the limit of spherical symmetry and neglecting the gas disk, $M_{*,\rm disk}+M_{*,\rm bulge}$ should be constant to recover the observed circular velocity; thus, $M_{*,\rm disk}$ and $M_{*,\rm bulge}$ are negatively correlated. This argument can be confirmed more quantitatively for the case of cylindrical symmetry as follows. For each value of $M_{*,\rm disk}$, we compute the unique baryon profile that gives the observed circular velocity and stellar and gas surface densities, fixing baryonic parameters as in the baseline analysis and holding the DM or ML parameters fixed. Specifically, for the ML case, we hold $\nu_\odot \equiv \nu_0+\nu_1\cdot a_{\rm N}(R_0) = 1.3$,\footnote{In the ML case, the circular velocity is only sensitive to $\nu_0$ and $\nu_1$ in the form of $\nu_{\odot}$.} approximately consistent with the median value from the scan (which is $\nu_\odot = 1.1$). This is in line with the intuition described in the previous sections: the ML scan picks out an acceleration enhancement that is close to unity. The results are plotted as the dashed lines in the central panel of \Fig{fig:Fits_1}, and we see that they agree appreciably well with the intuition outlined above.

The right panel of \Fig{fig:Fits_1} shows the correlation of $h_{*,R}$ with $M_{*,{\rm bulge}}$---a convolution of the results in the first two panels.  
As $h_{*,R}$ increases, the disk contribution to $a_{{\rm N},R}$ decreases, causing $M_{*,\rm bulge}$ to increase in order to maintain the constant $v_{c,\rm obs}$.  The DM and ML regions both show this general trend. The dashed lines representing the DM and ML estimates are obtained following the same procedure we used to get the estimates in the central panel.

\begin{figure}[t]
\centering
\includegraphics[width=.5\textwidth]{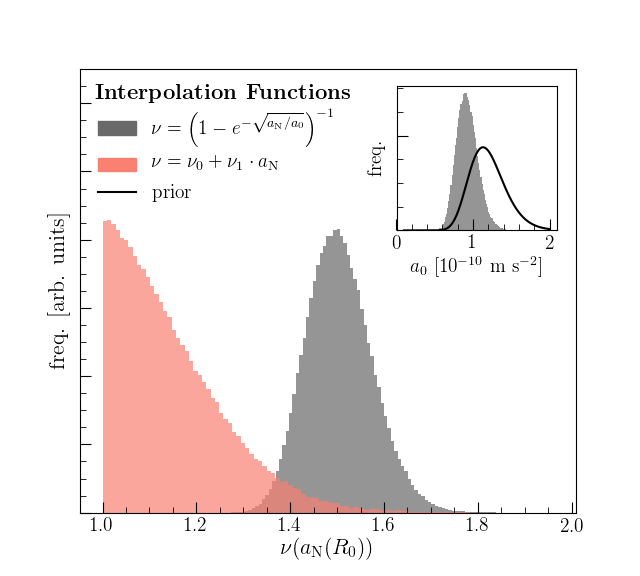}
\caption{Posterior distributions for the local acceleration amplification, $\nu(a_{N}(R_0))$, for two interpolation functions. The \textit{red} histogram corresponds to a Taylor expansion interpolation function and a prior of $\nu(a_{\rm N}) > 1$. The \textit{grey} histogram corresponds to the RAR interpolation function. The prior for the RAR function is plotted in the inset, together with the posterior of $a_0$, which is a free parameter of this scan. The posteriors of both interpolation functions push to low values of acceleration amplification, close to unity.}
\label{fig:MOND_Posterior}
\end{figure}

The open/filled black points in the right panel of \Fig{fig:Fits_1} represent two different measurements of the bulge mass: $M_{*, {\rm bulge,obs}} = 1.50 \pm 0.38$~$M_\odot$ from microlensing observations~\cite{Novatin:2007dd} (filled black circle) and  $M_{*, {\rm bulge,obs}} = 0.62 \pm 0.31$~$M_\odot$ (open black circle) from photometric observations~\cite{LopezCorredoira:2006na}.  Note that we have rescaled the latter to the value of $R_0$ used in this work~\cite{Licquia:2014rsa}. In both cases, the disk scale length is taken to be $h_{*, R, {\rm obs}} = 2.6\pm 0.5$~kpc. It is important to emphasize that we model the stellar disk using a single exponential profile, and thus $h_{*,R,\rm obs}$ is an effective value derived from a two-disk (thin and thick) formalism.  We perform a Monte Carlo procedure that properly combines the weighted contributions from the thin and thick disks.
For the thin disk, we adopt $h_{*,R,\rm obs}^t = 2.6 \pm 0.5$~\text{kpc} and $h_{*, z,\rm obs}^t = 300\pm50$~pc, and for the thick disk we adopt $h_{*,R, \rm obs}^T = 2.0\pm0.2$~kpc and $h_{*,z,\rm obs}^T = 900\pm180$~pc, as derived in \Ref{Bland-Hawthorn:2016aaa} from a review of the literature. We obtain the following estimates for the effective disk parameters:
\begin{equation}
  h_{*,R,\rm obs} = 2.6\pm 0.5~\text{kpc}\, \, \, \text{and} \, \, \, h_{*, z,{\rm obs}} = 310\pm50~\text{pc}\, .
  \label{eq:effective_disk}
\end{equation}
We note that $h_{*,z,\rm obs}$ is consistent with, although not precisely equal to, the benchmark value of $h_{*,z}=300\ \text{pc}$.

As mentioned above, in our final results, we will fix the scale length and bulge mass directly in the likelihood analysis. However, for illustrative purposes and because of the spread of values found in the literature (see \Ref{Bland-Hawthorn:2016aaa} and \Ref{Licquia:2014rsa} for a summary), we have also considered the case where these parameters are unconstrained in order to understand the general preference of the DM and ML models. Thus, the results presented in Fig.~\ref{fig:Fits_1} are quite general, and one can easily interpret them in the context of any specific measurement for the disk parameters. From this figure, we see that ML models prefer larger $M_{*,\rm bulge}$ and smaller $h_{*,R}$, as compared to DM, and have less overlap with the favored region indicated by the open/filled black points.  In order to understand this behavior, it is enlightening to study posterior distributions of a number of ML parameters.

Fig.~\ref{fig:MOND_Posterior} shows the posterior distributions of $\nu(a_{\rm N}(R_0))$ for two ML models. The red region corresponds to the baseline analysis with full freedom of the interpolation function, while the grey region corresponds to the analysis which enforces the RAR interpolation function (see Table~\ref{tab:nu_funcs}). The inset shows the prior and posterior distributions of $a_0$ for the RAR function.  Each model attempts to reduce the acceleration enhancement as much as is allowed within the adopted prior range.  For example,  the red region is pushed to the minimum allowed value of $\nu(a_{\rm N}(R_0))$ as allowed by the priors on $\nu_0$ and $\nu_1$ (\emph{i.e.}, $\nu_0 + \nu_1 a_{\rm N} > 1$).  The gray region is also forced to small values of $\nu(a_{\rm N}(R_0))$, with the location of the peak set by the fact that the $a_0$ posterior is peaked towards its lowest allowed value.  Because of the functional form of the interpolation function, as $a_0$ decreases, the local acceleration enhancement (the value of $\nu(a_{\rm N}(R_0))$) also decreases, \textit{i.e.}, accelerations become more Newtonian.  The gray region is still peaked above the red because it is associated with a specific interpolation function and preferred value of $a_0$.  This forces a model with the RAR interpolation function to significantly amplify local accelerations, making it challenging for this model to simultaneously explain radial and vertical accelerations.  On the other hand, the Tayler expansion function is able to explain the vertical accelerations by choosing $\nu(a_{\rm N}(R_0))\approx 1$, but then requires anomalous amounts of baryonic mass at the center of the Galaxy to explain the radial accelerations. 

\begin{figure}[t] 
\centering
\includegraphics[width=.45\textwidth]{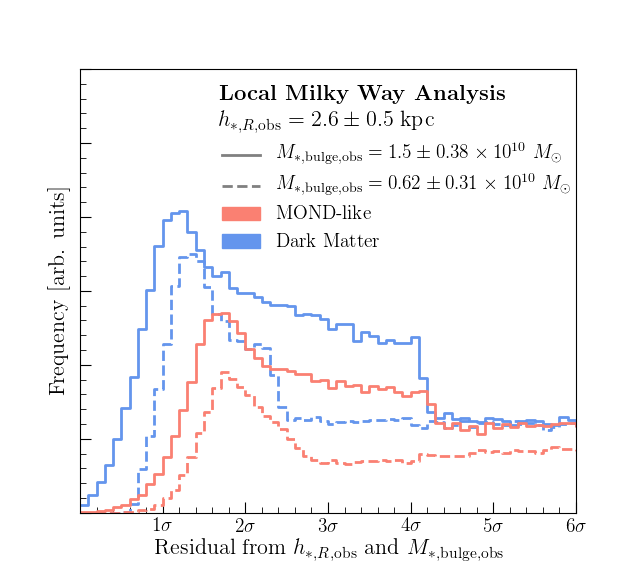}
\caption{Discrepancy of each walker step in the MCMC with measurements of the stellar disk scale length and bulge mass for the dark matter (\emph{blue lines}) and MOND-like (\emph{red lines}) models.  The discrepancy is calculated taking the scale length of \Eq{eq:effective_disk} and $M_{*, {\rm bulge, obs}} = 1.50 \pm 0.38$~$M_\odot$ from~\Ref{Novatin:2007dd} (\emph{solid lines}) and $M_{*, {\rm bulge, obs}} = 0.62 \pm 0.31$~$M_\odot$ from~\Ref{LopezCorredoira:2006na} (\emph{dashed lines}). In both cases, the distribution tends to peak at smaller discrepancies for the dark matter model than the MOND-like model. However, both distributions contain very long tails toward high discrepancies. Additionally, we note that the sharp drops in the distributions that can be seen near $\sim 4\sigma$ ($\sim 2.5\sigma$) in the solid (dashed) lines are due to the prior $M_{*,\rm bulge}>0$.}
\label{fig:Fits_3}
\end{figure}

\Fig{fig:Fits_3} plots the discrepancy of each step in the MCMC sample chain against the measured values $h_{*,R,\rm obs}$ and $M_{*,{\rm bulge, obs}}$;\footnote{We compute this discrepancy assuming that both measurements are uncorrelated and normally distributed. The discrepancy is then $\sqrt{\left(\frac{h_{*,R}-h_{*,R,\rm obs}}{\delta h_{*,R,\rm obs}}\right)^2 + \left(\frac{M_{*,\rm bulge}-M_{*,\rm bulge,obs}}{\delta M_{*,\rm bulge,obs}}\right)^2}$, where $\delta h_{*,R,\rm obs}~(\delta M_{*,\rm bulge,obs})$ is the measurement error of the scale length (bulge mass).} the solid (dashed) lines correspond to the larger (smaller) bulge mass measurement.  This result is shown only for the baseline DM and ML scans.  
The majority of the walker steps fall within $\sim 1\sigma$ of the $h_{*,R}$ observation for the DM model, while the corresponding distribution for the ML case is peaked around $\sim2\sigma$ discrepancy. This figure is an alternative way to represent the results from the right panel of Fig.~\ref{fig:Fits_1}, and highlights the fact that the DM model is more consistent with observations of the baryonic profile.

We can also use the Bayesian Information Criterion (BIC)~\cite{10.2307/2291091} to compare the DM and ML models, defining
\begin{equation}
\Delta \text{BIC} = \text{BIC}_{\rm ML} - \text{BIC}_{\rm DM} \, .
\label{eq:deltaBIC}
\end{equation}
The larger the value of $\Delta \text{BIC}$, the more strongly disfavored the ML model is relative to the DM model. To obtain these values, we run the scans imposing additional constraints on the parameters.  If we take the scale length to be that of \Eq{eq:effective_disk} and the bulge mass to be $1.50\pm0.38 \times 10^{10}$~($0.62\pm0.31 \times10^{10}$)~$M_\odot$, then $\Delta \text{BIC} = 4.1$~(4.5) for the baseline scan where the local interpolation function is constrained only to be larger than 1.  These values suggest a positive, but not strong, preference for the DM model. Notice that the $\Delta \text{BIC}$ value increases when the bulge mass is smaller.  We have also calculated the $\Delta \text{BIC}$ for scans with a number of interpolation functions as detailed in Table~\ref{tab:nu_funcs}. The Taylor expansion function with $\nu(a_{\rm N}) > 1$ corresponds to the results given above. The other scans result in larger $\Delta \text{BIC}$ values (given for a bulge mass of $1.50\pm0.38\times10^{10}$~$M_{\odot}$) as follows. For the Taylor expansion function with $\nu(a_{\rm N}) > 1.3$, $\Delta \text{BIC} = 7.5$.  For the RAR function, $\Delta \text{BIC} = 10.4$.  For the Simple function, $\Delta \text{BIC} = 9.6$.  For the Standard function, $\Delta \text{BIC} = 4.8$. All these results more strongly favor a DM model. The underlying reason being that these functions enforce a local acceleration enhancement substantially larger than unity and the tension described in the previous sections becomes more apparent.

Finally, in line with the reasoning described in Sec.~\ref{sec: framework}, we have performed a scan with a DM halo that is slightly prolate. In this case, the ratio of $a_{{\rm DM},z}/a_{{\rm DM},R}$ is even more suppressed relative to a ML model. We take the halo to have the functional form described in Ref.~\cite{Bovy:2016chl}, which is equivalent to \Eq{eq:rho_DM} with the replacement
\beq
r^2 \rightarrow R^2 + \left(\frac{z}{c/a}\right)^2.
\eeq
We take the best-fit axis ratio from that study, namely $c/a=1.05$. In this scan, we impose the constraints $M_{*,{\rm bulge}} = 1.50\pm0.38 \times 10^{10}$~$M_\odot$ and $h_{*,R} = 2.6\pm 0.5$~kpc. This model is preferred over a ML model with $\Delta \text{BIC} = 4.2$. We have verified that allowing $c/a$ to scan to larger values increases this preference.

\subsection{Kinematic Comparison of the Models}

\begin{figure}[t] 
\centering
\includegraphics[width=.5\textwidth]{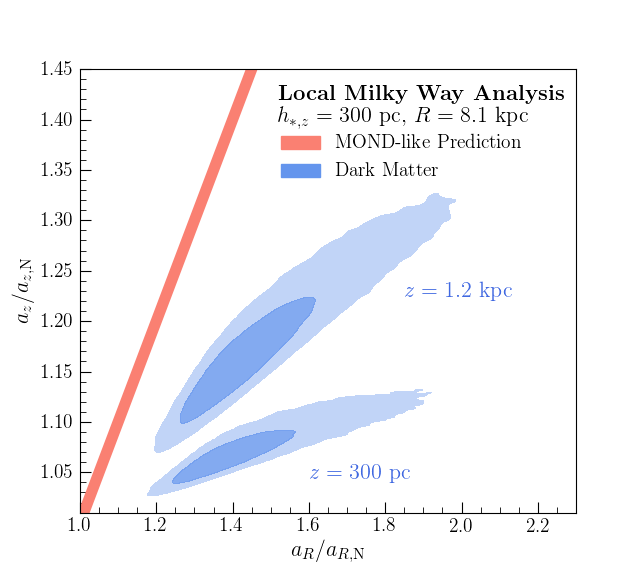}
\caption{Correlation of observables that highlight the main kinematic difference between the MOND-like and dark matter scenarios. The axes correspond to the ratio of the observed acceleration to the Newtonian (baryon-only) acceleration in the vertical and radial directions.  The accelerations are evaluated at the Solar radius, $\Rsun$.  Models satisfying \Eq{eq:nu_aobs} identically predict that observations must satisfy $a_z/a_{\rm{N},z}=a_R/a_{\rm{N},R}$ (\textit{red line}), which is not satisfied by a spherically symmetric dark matter halo.  The {\it blue shaded regions} correspond to 68 and 95\% containment for the dark matter analysis at reference positions $z=300$ and $1200$~pc.}
\label{fig:Fits_2}
\end{figure}

As discussed in \Sec{sec: framework}, the primary tension between ML models and MW observables lies in the inability to simultaneously reproduce the radial and vertical accelerations of stars. As a test of this, we consider the correlation between the ratios $a_{z}/a_{\text{N}, z}$ and $a_{R}/a_{\text{N}, R}$, evaluated at $R=\Rsun$ and at $z= 300$~and 1200~pc. Fig.~\ref{fig:Fits_2} plots these two ratios for the models under consideration.  The ML model directly relates each of these ratios to the interpolating function, $\nu(a_{\rm N}/a_0)$, so they are always equivalent.  In other words, in a ML model, the ratio of the observed and the Newtonian accelerations is independent of direction. The ML model prediction is simply the solid red line in \Fig{fig:Fits_2}.

The DM case is comparatively more interesting.  The difference between the DM confidence regions and the ML expectation highlights the main kinematic difference between the two models, as anticipated in the discussion of \Sec{sec: framework}.
If the DM best-fit region were to partially overlap with the ML line, then the models would be kinematically indistinguishable in a certain region of parameter space.
We instead see that there is always a sizable difference between the two cases, for all values of $z$ between 300--1200~pc. We also note that since the slope of the DM correlation should be set by $z/R_0$, the DM and ML kinematics become more and more similar as $z$ is increased. This is because the disk potential looks more spherical as $z$ increases. In the limit of a spherically symmetric baryonic distribution, we expect ML models and DM to be indistinguishable, at least in the sense described in \Sec{sec: framework}.

\subsection{Dark Matter and MOND-like Parameters}

\begin{figure*}[htbp] 
\centering
\includegraphics[width=.49\textwidth]{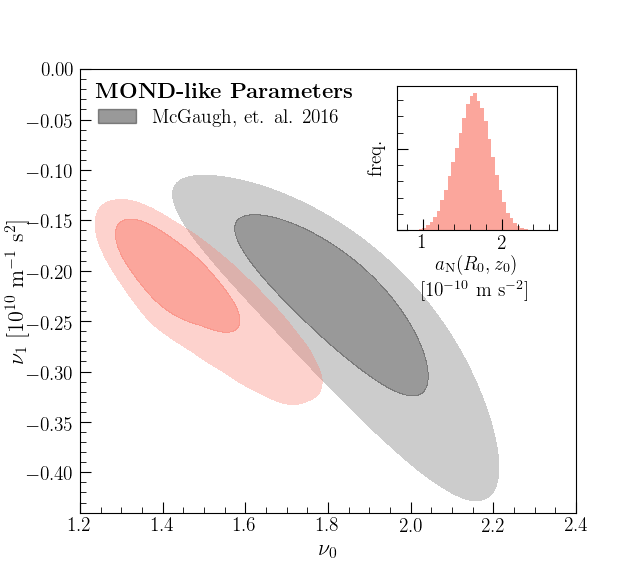}\includegraphics[width=.49\textwidth]{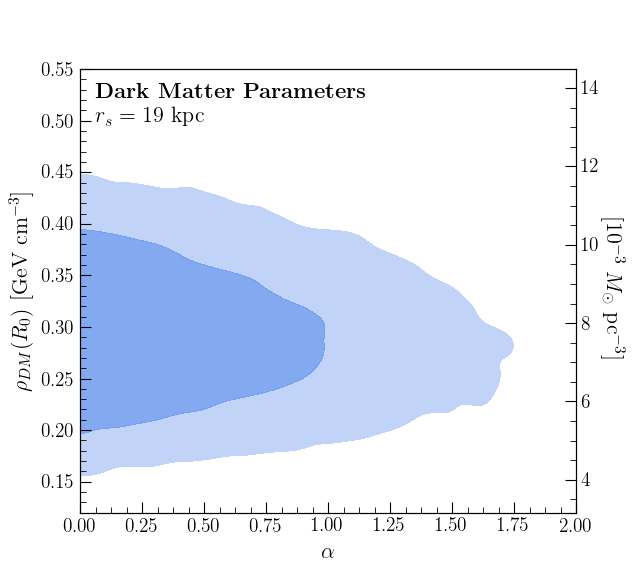}
\caption{Marginalized two-parameter correlations for the parameters of the dark matter and MOND-like models analyzed in this study. Shaded regions correspond to 68\% and 95\% containment. {\bf Left panel}: Parameters of the MOND-like model. The analysis very tightly constrains the model-independent values $\nu_0$ and $\nu_1$ (\textit{red shaded region}). Given any interpolation function, its consistency with the results of this analysis can be tested by computing $\nu_0$ and $\nu_1$ from \Eq{eq:invmu}, taking $a_{\rm N,ref}$ to be distributed according to the inset distribution shown, and then comparing with the red shaded region. As an example we perform this procedure for the RAR interpolation function defined in Table~\ref{tab:nu_funcs}, which is shown by the \textit{gray shaded region}. Our analysis disfavors this interpolation function. {\bf Right panel}: Parameters of the dark matter model using a modified NFW profile with inner profile slope parameter, $\alpha$, and local density, $\rho_{\rm DM}(\Rsun)$. The fully marginalized value of the local density, $\rho_{\rm DM}(\Rsun) = 0.29\pm0.06\ \mathrm{GeV}\ \mathrm{cm}^{-2}$, is consistent with previous measurements. The posterior values of $\alpha$ are consistent with a standard cusped NFW profile as well as a more cored profile with $\alpha$ close to zero. In particular, we find that $\alpha < 1.1$ at 90\% confidence.}
\label{fig:MOND_DM_Params}
\end{figure*}

In addition to allowing us to compare the DM and ML models, our Bayesian analysis framework allows us to recover the best-fit parameters specific to each. For example, the right panel of Fig.~\ref{fig:MOND_DM_Params} shows the 68\% and 95\% containment region for the local DM density, $\rho_\text{DM} (R_0)$, and the inner slope of the NFW profile, $\alpha$. Although $\rho_{\rm DM}(\Rsun)$ is not an independent free parameter of our framework, it is easily computable from \Eq{eq:rho_DM}. We find that the marginalized value for the local DM density is $\rho_{\rm DM}(\Rsun) = 0.29\pm0.06\ \mathrm{GeV}\ \mathrm{cm}^{-2}$, consistent with previous measurements summarized in~\Ref{Read:2014qva}. In particular, we can compare our results to those of \Ref{Zhang:2012rsb}, whose K-dwarf number densities and vertical dispersions are used here (see \Sec{sec: constraints}). The two studies differ in several important analysis details---for example, we vary over the full baryonic profile and assume an NFW distribution.  However, the final results are similar, with \Ref{Zhang:2012rsb} finding a local DM density of $0.25\pm0.09$~GeV~cm$^{-3}$.
We also constrain the fully marginalized value of the inner slope of the NFW profile to be $\alpha < 1.1$ at 90\% confidence. This result, which is consistent with previous dynamical studies~\cite{Bovy:2013raa}, is not constraining enough to distinguish between a cored ($\alpha\approx0$) and cusped profile ($\alpha\approx1$).  

The left panel of \Fig{fig:MOND_DM_Params} shows the analogous containment for the ML parameters, $\nu_0$ and $\nu_1$, in the red shaded region, for the baseline scan with $\nu(a_{\rm N})>1$. We find that the fully marginalized values are $\nu_0 = 1.4\pm 0.1$ and $\nu_1 = -0.21^{+0.03}_{-0.05}\times 10^{10}\ \mathrm{s}^2\ \mathrm{m}^{-1}$. Because $\nu_0$ and $\nu_1$ are a model-independent parametrization of the interpolation function, these values can be related to any specific interpolation function. As an example, we consider the RAR interpolation function of \Ref{McGaugh:2016leg}, which is given in Table~\ref{tab:nu_funcs} with $a_0=1.20\pm0.24\times 10^{-10}\ \mathrm{m}\ \mathrm{s}^{-2}$ (note that we have now chosen the specific central value and errors of $a_0$ which is the best fit to the RAR data). In order to map the interpolation function onto the $\nu_0$ and $\nu_1$ parameters, the expansion parameter of \Eq{eq:invmu}, $a_{\rm N,ref}$, must be specified. Although in principle $a_{\rm N,ref}$ can be evaluated at any point within the local region of interest, it will differ very little from $a_{\rm N}(\Rsun,\zsun)$; we therefore approximate $a_{\rm N,ref}\approx a_{\rm N}(\Rsun,\zsun)$.\footnote{We have verified that the gray region in the left panel of Fig.~\ref{fig:MOND_DM_Params} does not change appreciably when $a_{\rm N, ref}$ is evaluated within $|R-\Rsun| \lesssim 500\ \mathrm{pc}$ and $|z|\lesssim 2\ \mathrm{kpc}$.}  We then perform the following Monte Carlo procedure to estimate the values of $\nu_0$ and $\nu_1$ that correspond to the RAR interpolation function with this value of $a_0$: for each step in the MCMC sample chain for the ML model, we compute $a_{\rm N}(\Rsun,\zsun)$, generate a random value of $a_0$ that is normally distributed with mean and standard deviation as specified by \Ref{McGaugh:2016leg}, and then compute $\nu_0$ and $\nu_1$ using Eq.~\eqref{eq:invmu}. The gray shaded regions in the left panel of \Fig{fig:MOND_DM_Params} indicate 68\% and 95\% containment for the recovered values of $\nu_0$ and $\nu_1$. We note that the region of maximum posterior probability for $\nu_0$ and $\nu_1$ in our analysis is beyond the gray shaded countours, indicating that our analysis disfavors the interpolation function of \Ref{McGaugh:2016leg} together with its best-fit values for $a_0$.  We note, however, that this result is specific to this specific interpolation function.  A different function could simultaneously fit the RAR and be consistent with our results.

We stress that our treatment of the interpolation function from \Ref{McGaugh:2016leg} is purely demonstrative.
Given any interpolation function and value of $a_0$, one can check its consistency against our results.  To this end, we have included the fully marginalized posterior distribution for $a_{\rm N}(\Rsun,\zsun)$ in the inset of the left panel of  \Fig{fig:MOND_DM_Params}.
The distribution is well-approximated by a Gaussian with mean $1.65\times 10^{-10}\ \mathrm{m}\ \mathrm{s}^{-2}$ and standard deviation $0.22\times 10^{-10}\ \mathrm{m}\ \mathrm{s}^{-2}$.
Although we used the full MCMC sample chain to construct the gray region in \Fig{fig:MOND_DM_Params}, we have verified that approximating the $a_{\rm N}(\Rsun,\zsun)$ distribution with the aforementioned Gaussian does not significantly affect the results and should serve as a sufficient estimator for one who seeks to perform this procedure without access to the  complete MCMC sample chain.

\subsection{Systematic Checks}
\label{sec:uncertainties}

Lastly, we explore the effects of varying the baseline analysis by repeating the MCMC scan for different values of the fixed parameters. To assess the overall impact on the final conclusions, we impose the constraints $h_{*,R, {\rm obs}} = 2.6\pm0.5$~kpc and $M_{*,{\rm bulge, obs}} = 1.50\pm0.38\times10^{10}$~$M_\odot$. We quote the $\Delta \text{BIC}$ as defined in \Eq{eq:deltaBIC}, and compare to the baseline value of 4.1, for the  following variations:
\begin{itemize}
\item fix the bulge scale radius to $r_{*,\rm bulge} = 2$~kpc, rather than $600$~pc.   ($\Delta \text{BIC} =3.7$)
\item fix the NFW scale radius to $r_s = 25$~kpc, rather than 19~kpc.  ($\Delta \text{BIC} =  3.9$)
  \item fix the gas scale length to $h_{g, R} = 4.5$~kpc~\cite{2003ApJ...588..805K,2016A&A...593A.108B}, rather than requiring $h_{g, R} = 2 \, h_{*, R}$~kpc ($\Delta \text{BIC} = 4.6$).  For comparison, the median values of the stellar scale length for the ML (DM) baseline study is $h_{*, R} = 2.41~(2.54)$~kpc, corresponding to a median gas scale length of $h_{g, R} = 4.82~(5.08)$~kpc.
\item fix the gas scale height to $h_{g, z} = 180$~pc, rather than 130~pc. ($\Delta \text{BIC} = 4.2$)
\item change the stellar disk scale height to $h_{*,z } = 200, 400, 600$~pc, compared to 300~pc.  ($\Delta \text{BIC} = 4.1, 3.5, 2.9$ respectively)
\item allow $h_{*,z}$ to vary as a free parameter of the scan and apply the constraint that $h_{*, z,{\rm obs}} = 310 \pm 50$ pc. ($\Delta \text{BIC} = 4.1$)
\item use logarithmic, rather than linear, priors for all parameters in the scan.  ($\Delta \text{BIC} = 3.4$)
\end{itemize}
In general, we find that the conclusions are not strongly sensitive to the variations listed above, with the results always providing positive evidence in favor of DM.  Of the parameters that we varied over, the analysis is perhaps the most sensitive to the stellar disk scale height.  To understand this behavior, we observe that increasing the disk scale height, while fixing the local stellar surface density, decreases the stellar density close to the midplane. As a result, the vertical velocity dispersion decreases as well, forcing a larger value of $\nu_\odot$ in order for a ML model to match observations. Because the radial acceleration obeys the scaling $a_R\simeq \nu_\odot a_{R, N}$, a larger $\nu_\odot$ is compensated by a smaller $a_{R, N}$ to reproduce the observed circular velocity.  This, in turn, implies a smaller stellar bulge mass and/or larger stellar scale radius, in better agreement with observations. 

As an additional variation, we replace the effective stellar disk with two double exponential profiles---one each for the thin and thick disk---to verify that the use of a more complicated density profile does not dramatically affect the results.  For simplicity, we fix the thick disk parameters and the thin disk scale height using the best-fit values from the SDSS study in \Ref{Juric:2005zr}, and we keep the scale length of the thin disk as a free parameter in the MCMC scan. The gas disk parameters are treated in the same way as in the baseline study.  In this case, we find $\Delta \text{BIC} = 4.1$.

An important assumption entering into the Jeans analysis described in Sec.~\ref{sec: constraints}---specifically \Eq{eq:sigz}---is that the tracer stars are in steady state.  This assumption is challenged by recent evidence of waves in nearby disk stars, possibly caused by the passage of a dwarf galaxy such as Sagittarius through the disk~\cite{Widrow:2012wu, 2013MNRAS.436..101W,Gomez:2012rd,2018Natur.561..360A, 2019MNRAS.482.1417B}.  Such perturbations can lead to $\mathcal{O}(10\%)$ changes to the vertical force~\cite{2017MNRAS.464.3775B}.   We have performed systematic checks to see how robust our results are to the uncertainties in the constraints used.  For example, we repeated the baseline MCMC scan, (arbitrarily) doubling the errors on the tracer number density and vertical dispersion data from~\Ref{Zhang:2012rsb}.  While we still find positive evidence in favor of DM ($\Delta \text{BIC} = 3.1$), the preference is smaller.  This is expected because larger errors would reduce the tension between the ML prediction and the vertical acceleration data.  As an additional test, we (artificially) double the errors on the surface density constraint taken from~\Ref{McKee:2015hwa}; in this case, $\Delta \text{BIC} = 4.1$, and is unchanged.

\section{Conclusions}
\label{sec: conclusions}

In this work, we tested the consistency of DM and models which resemble MOND on galactic scales with local MW observables. Our goal was to compare the models on galactic scales, where ML models have had their greatest success. As a concrete example, we focused on scalar enhancements to Newtonian gravity, where the absolute magnitude, but not the direction, of the Newtonian acceleration is altered. In this case, the direction of the total acceleration vector is set by the baryonic mass distribution.  This is fundamentally different than the DM expectation; in the presence of a spherical halo, the total acceleration vector necessarily points towards its center. 

Such phenomenology is generically predicted by models which reduce to a ML force on galactic scales. One example is Superfluid DM~\cite{Berezhiani:2015bqa, Famaey:2017xou, Berezhiani:2017tth}, where the ML force arises as an emergent force in a superfluid phase of DM. In that model, there are additional complications arising from the existence of a small DM density together with the ML force. While we postpone the full treatment of these models to future work, we expect a qualitative similarity to the current study where we consider the case of a ML force alone.  In general, we find that scalar enhancements to gravity over-predict the vertical acceleration of nearby stars when trying to simultaneously explain the observed radial acceleration. This provides a crucial handle for testing these models on galactic scales.

We have performed a Bayesian likelihood analysis to compare DM to ML models.  The results depend sensitively on the baryonic mass distribution, which we describe using a stellar disk, a gas disk, and a stellar bulge.  We varied over the parameters of the baryonic profile, imposing minimally restrictive priors.  For the DM scenario, the halo distribution was assumed to follow a spherical NFW profile. For the ML scenario, we focused on the class of models where the baryonic acceleration is enhanced by a scalar function.   For our baseline analysis, we parametrized this scalar interpolation function in a model-independent manner by considering its linear approximation near the Solar position. This approach allowed us to evaluate the preference for DM, independent of the details of the interpolation function.  The results are generic and conservative, as the parametrization has total freedom to enhance accelerations as much or as little as required by the data (including for example no enhancement at all).  In this case, we find positive, but not strong, preference for DM relative to ML models.  However,  when we compare DM to ML models with specific interpolation functions, we find strong tension with data.

The free model parameters were constrained using local measurements of the baryonic surface density, rotational velocity, and vertical dispersion of tracer stars. We note that, on general grounds, the non-linear nature of ML models might result in an acceleration field that is not parallel to the Newtonian prediction. However, as is demonstrated in \App{sec:non_linear_effects_in_mond}, we expect any deviations from alignment with the Newtonian acceleration to be small. 

Our findings can be summarized as follows.  In comparison to DM, ML models typically prefer a smaller scale length for the stellar disk and a larger stellar bulge mass, in tension with current measurements.  The underlying reason for this is that MW data seem to prefer a model which is able to simultaneously enhance radial acceleration with little enhancement to vertical accelerations. ML models are generically unable to provide such phenomenology unless they invoke anomalously large amounts of baryonic mass at the Galactic Center. For our baseline analysis and taking an observed scale length of $h_{*, R, {\rm obs}} = 2.6\pm0.5$~kpc and bulge mass of $M_{*, {\rm bulge}, {\rm obs}} = 1.50\pm0.38 \times 10^{10}$ ($0.62\pm0.31 \times10^{10}$)~$M_\odot$, we find that $\Delta \text{BIC} = 4.1$~(4.5), favoring the DM model. These conclusions are robust to systematic variations of the analysis.  We also find that a slightly prolate DM halo is also consistent with the data.  When the analysis is done with any specific interpolation function, forcing acceleration enhancements to be considerably larger than unity, the $\Delta \text{BIC}$ grows to values in the range 4.8--10.4, depending on the specific functional form and its prior. 
 
The tension that we have observed for ML models versus DM will be further clarified as measurements of the baryonic density profile continue to improve.  As we have seen, the disk scale length and bulge mass play a particularly important role in defining this tension, so reducing the measurement uncertainty on these parameters has the potential to strengthen the conclusions.  Additionally, we note that the measurements of the MW rotation curve are systematics-dominated and that improvements in the measurement of the local circular velocity would also strengthen the power of the analysis by tightening the contour bands in \Fig{fig:Fits_1}.  As a simple demonstration of this, we have rerun the analysis using only statistical errors on these quantities, ignoring the quoted systematic uncertainties.  This increases the preference for the DM model, with $\Delta\rm{BIC}=7.1$.  

This initial study does not take full advantage of the \emph{Gaia} dataset~\cite{2018A&A...616A...1G}.  Indeed, the only input that we have used from \emph{Gaia} is the recent update to the rotation curve from \Ref{2018arXiv181009466E}.  As the characterization of local MW dynamics continues to improve with \emph{Gaia}, one can revisit the analysis proposed in this work to obtain an even more definitive test of the models. 

\section{Acknowledgements}  
We thank J.~Bovy, J.~Herzog-Arbeitman, D.~Hogg, G.~Kribs, S.~McGaugh, H.~Walter-Rix, H.~Verlinde, and L.~Zhang for useful conversations.  This research made use of the \texttt{Astropy}~\cite{2013A&A...558A..33A}, \texttt{IPython}~\cite{PER-GRA:2007}, \texttt{matplotlib}~\cite{Hunter:2007}, \texttt{numpy}~\cite{Oliphant:2015:GN:2886196}, \texttt{galpy}~\cite{2015ApJS..216...29B}, \texttt{corner}~\cite{Foreman-Mackey:corner}, and \texttt{emcee}~\cite{2013PASP..125..306F} software packages. ML is supported by the DOE under contract DESC0007968 and the Cottrell Scholar Program through the Research Corporation for Science Advancement.  MM is supported by the DOE under contract DESC0007968. NJO is supported by the Azrieli Foundation Fellows program.
\twocolumngrid

\begin{appendix}
\section{Non-Linear Effects in MOND and MOND-like models} 
\label{sec:non_linear_effects_in_mond}

\subsection{The Divergenceless Field} 
\label{sub:S_Field}

Any analysis of the type presented in this study critically assumes some correlation between the Newtonian acceleration that is expected from the baryonic distribution and the observed acceleration. Because ML models typically invoke a non-linear response to the distribution of baryonic matter, any simple relation between Newtonian and observed acceleration may hold only approximately. If this is the case, then one must first ensure that any corrections do not significantly affect the analysis. For our analysis, we have assumed Eq.~(\ref{eq:nu_aobs}) to be true, and we now describe the possible corrections to this and why we believe them to be small.

Eq.~(\ref{eq:nu_aobs}) is in fact only a special case of the correlation between the observed and Newtonian acceleration, and is known to violate standard conservation laws~\cite{Milgrom:1983ca}. To circumvent this, it is therefore required to define a scalar field whose gradient is the acceleration field, in analogy with Newtonian gravity.  A simple realization is provided in Ref.~\cite{Milgrom:2009ee} as
\begin{equation}\label{eq:lap_qumond}
	\boldsymbol{a}=-\boldsymbol{\nabla}\Phi\;\;,\;\;\nabla^2\Phi=-\boldsymbol{\nabla}\cdot\left(\nu\left(\frac{a_N}{a_0}\right)\boldsymbol{a_N}\right).
\end{equation} 
When the Newtonian acceleration is a function of a single coordinate, it can be shown that the above equation reduces to Eq.~(\ref{eq:nu_aobs}) using the divergence theorem~\cite{Brada:1994pk,Bekenstein:1984tv}. More generally, however, \Eq{eq:lap_qumond} holds up to a divergence-less vector field, $\boldsymbol{S}\equiv\boldsymbol{\nabla}\times \boldsymbol{h}$, such that\footnote{Strictly speaking, Eqs.~(\ref{eq:lap_qumond},\ref{eq:mu_aobs}) correspond to a version of MOND known as Quasi-Linear MOND (QuMOND)~\cite{Milgrom:2009ee}. However, both the standard formulation of MOND and QuMOND are equally motivated, and they reduce to each other exactly in the limit where $\boldsymbol{S}=0$.}
\beq
\boldsymbol{a} = \nu\left(\frac{a_{\rm N}}{a_0}\right) \boldsymbol{a}_{\rm N} + \boldsymbol{S} \, .
\label{eq:mu_aobs}
\eeq
The MW baryonic profile is disk-like, and even in the limit of a perfect disk, two coordinates are required to describe it (namely $R$ and $z$).  Thus, in principle, $\boldsymbol{S}$ need not be zero and other considerations must be used to ensure that any effects arising from a non-zero $\boldsymbol{S}$ field are negligibly small.

We are interested in ensuring that we can safely assume that $\boldsymbol{S} \ll \nu\left(a_{\rm N}/a_0\right) \boldsymbol{a}_{\rm N}$ in the region of study used in this work.  Following the arguments of Ref.~\cite{Brada:1994pk}, since $\boldsymbol{S}$ is both divergenceless and irrotational, the following self-consistency condition relates the requirement of vanishing $\boldsymbol{S}$ with a requirement on the Newtonian potential:
\begin{equation}
\boldsymbol{S}=0\;\;\leftrightarrow\;\; \boldsymbol{\nabla}| \boldsymbol{\nabla} \Phi_{\rm N} | \times \boldsymbol{\nabla}\Phi_{\rm N} = 0 \, .
\label{eq:Szero_PhiN}
\end{equation}
This equation is satisfied when $\boldsymbol{\nabla}\Phi_{\rm N}=0$ or when $\boldsymbol{\nabla}|\boldsymbol{\nabla} \Phi_{\rm N}|$ is parallel to $\boldsymbol{\nabla}\Phi_{\rm N}$, and can be verified explicitly for any Newtonian potential. Specifically, the same study found this requirement to hold extremely well for any baryonic profile with a radial exponentially decreasing surface density profile. We have repeated the same analysis assuming a baryonic profile that consists of a stellar disk ($M_{*,{\rm disk}}=3.5\pm1\times10^{10}$ $M_{\odot}$, $h_{*, R}=2.6\pm0.5$~kpc, $h_{*, z}=300\pm50$~pc), a gas disk ($M_{g,{\rm disk}}=1\pm0.4\times10^{10}$~$M_{\odot}$, $h_{g,R}=5.2\pm1$~kpc, $h_{g, z}=130\pm50$~pc) and a Hernquist stellar bulge ($M_{*,{\rm bulge}}=1\pm0.4\times10^{10}$~$M_{\odot}$), finding that any deviation from Eq.~(\ref{eq:Szero_PhiN}) cannot be distinguished to within measurement errors of $|\boldsymbol{\nabla} \Phi_{\rm N}|$. The values of the baryonic profile parameters used here are motivated by measurements of the bulge mass~\cite{Licquia:2014rsa}, the stellar~\cite{Bland-Hawthorn:2016aaa} and gas~\cite{2008gady.book.....B,Bovy:2013raa} disk parameters ($M_{g,{\rm disk}}$ is the total mass of the gas disk) as well as local surface density measurements~\cite{McKee:2015hwa}. We have also varied over these parameters, and find that this result is not sensitive to any $\mathcal{O}(1)$ changes to their values. The advantage of this type of consideration is that it does not require any knowledge of $\nu(a_{\rm N}/a_0)$ and is therefore robust. The disadvantage is that it is not straightforward to quantify the size of $\boldsymbol{S}$ given that Eq.~(\ref{eq:Szero_PhiN}) is only approximately satisfied.

\subsection{Mass Perturbations} 
\label{sub:effect_of_a_small_perturbation}

An additional complication arises from the fact that the MW baryonic density profile is not perfectly smooth. Because of the non-linear nature of MOND and ML models, small perturbations to a smooth matter distribution can potentially cause extremely large effects. This is related to the previously studied External Field Effect~\cite{Bekenstein:1984tv,Milgrom:1983ca,Famaey:2011kh,Haghi:2009mg,Hees:2015bna}. Thus, if one is to model the MW with a smooth cylindrical density profile, it is of the utmost importance to ensure that small perturbations do not cause large effects in the region of study.

Near the Sun's position in the MW disk, there are a number of known stellar overdensities  \cite{Juric:2005zr}. These regions explicitly break assumptions regarding cylindrical symmetry and the smoothness of the baryonic density distribution. In Newtonian gravity, the acceleration field is dictated by the Poisson equation, whose linearity enables one to very simply determine when a perturbation can be neglected. The non-linear nature of ML models requires a more subtle treatment since, as will be shown below, the effects of a perturbation can be felt over larger distances.

\begin{figure}[t]
  \begin{center}
    \includegraphics[width=.48\textwidth]{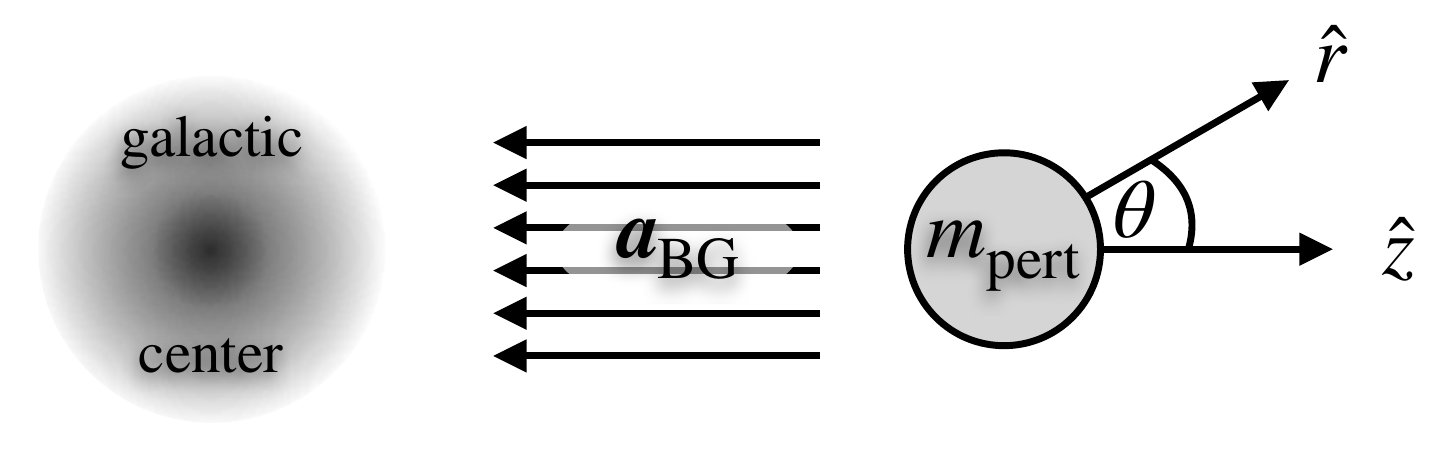}
    \caption{\label{fig:pert_diagram} A schematic diagram illustrating the geometry described in the text. $m_{\rm pert}$ is a small mass perturbation to a smooth background acceleration field, $\boldsymbol{a}_{\rm BG}$, which is approximately constant in the vicinity of the perturbation. Choosing the coordinate system as depicted in the diagram, the MONDian potential around the perturbation is given by \Eq{eq:pert_qumond}.}
  \end{center}
\end{figure}

To quantify this effect, the above situation can be modeled as follows. Consider a point mass, $m_{\rm pert}$, at a position within the disk, but sufficiently far from the Galactic Center such that the background Galactic acceleration field (in the vicinity of the point mass), $\boldsymbol{a}_{\rm BG}$, can be treated as constant. The point mass creates its own acceleration field which depends on $\boldsymbol{a}_{\rm BG}$ because of the non-linear behavior of the model. We are interested in quantifying the size of this field and comparing it to the local observed acceleration, $\boldsymbol{a}_{\rm loc}$, which can be directly inferred observationally, \emph{e.g.}, from the local value of the rotation curve,
\begin{equation}
a_{\rm loc}=\frac{v_c^2}{R_0} \approx 2\cdot10^{-10} \text{ m/s}^2 \, .
\end{equation}
Setting the coordinate origin to the position of the point mass and aligning the $\hat{z}$ direction with $-\boldsymbol{a}_{\rm BG}$ (see Fig.~\ref{fig:pert_diagram}), the Newtonian potential around the point mass can be written as
\begin{equation}\label{eq:pert_newton}
\Phi_{\rm N} = -a_{\rm N,BG} \, z - \frac{G \, m_{\rm pert}}{r},
\end{equation}
where $a_{\rm N,BG}$ is the Newtonian background field that relates to the observed background field through $\nu(a_{\rm N,BG}/a_0) \, a_{\rm N,BG} = a_{\rm BG}$. To leading order in $m_{\rm pert}$, the perturbed field (background and perturbation) is given by
\begin{equation}\label{eq:pert_qumond}
\Phi=-a_{\rm BG}\,z-\frac{G \, m_{\rm pert}}{r}\, \nu\left( \frac{a_{\rm N,BG}}{a_0} \right) \left[ 1+\frac{\gamma_{\rm BG}}{2}\sin^2 \theta \right] \, ,
\end{equation}
where $\gamma_{\rm BG} \equiv \partial \log \nu / \partial \log{[a_{\rm N}/a_0]}$ evaluated at $a_{\rm N,BG}/a_0$~\cite{Milgrom:2009ee}. The result of Eq.~\eqref{eq:pert_qumond} allows one to determine the maximal range of influence of the perturbation by demanding that its effect be much smaller than $a_{\rm loc}$. This requires that the distance, $D$, from the perturbation satisfies
\begin{equation}\label{eq:pert_range}
D\gg\sqrt{\nu\left( \frac{a_{\rm N,BG}}{a_0} \right) \frac{G \, m_{\rm pert}}{a_{\rm loc}}} \,.
\end{equation}
Note that $D$ is always larger than the equivalent distance for the Newtonian case where $\nu=1$.

There are a number of known stellar overdensities within $\mathcal{O}(1)$~kpc of the Sun. The most prominent of these are localized at $(R,Z)\sim(6.5,1.5)\,\rm {kpc}$ and $\sim(9.5,0.8)\,\rm kpc$, with densities of approximately twice that of their average surroundings and sizes of order~kpc$^3$---see, \emph{e.g.}, Fig. 27 of~\Ref{Juric:2005zr}.  Comparing to the smooth density profile model found by these authors at the locations of these perturbations, we estimate the total mass of each overdensity to be  $m_{\rm pert}\lesssim 10^7\,M_{\odot}$.  To safely neglect the effects of these overdensities, the distance from the mass perturbation must be larger than
\beq
D \gg 0.1 \text{ kpc} \cdot \left[\nu\left( \frac{a_{\rm N,BG}}{a_0} \right) \cdot \frac{m_{\rm pert}}{10^7\,M_{\odot}} \cdot \frac{2\cdot10^{-10}\text{ m/s}^2}{a_{\rm loc}}\right]^{1/2},
\eeq
which is automatically satisfied for any interpolation function whose value in the vicinity of the perturbation is $\nu\approx1$.

\section{MCMC details and convergence} 
\label{sec:mcmc_convergence}

\begin{figure*}[h]
  \begin{center}
    \includegraphics[width=.99\textwidth]{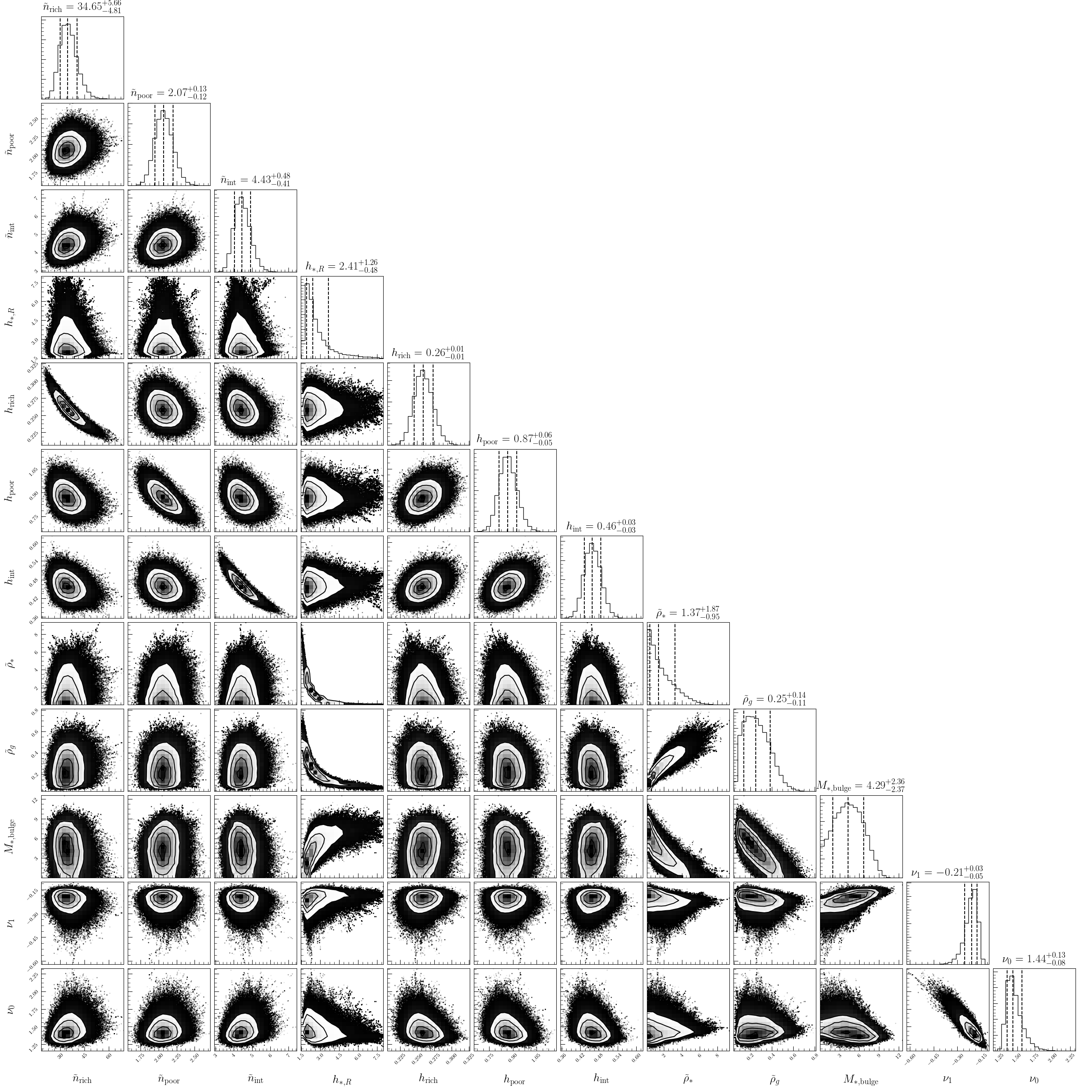}
    \caption{\label{fig:mond_corner} The posterior distributions and two-parameter correlations for the MOND-like model parameters.}
  \end{center}
\end{figure*}

\begin{figure*}[h]
  \begin{center}
    \includegraphics[width=.99\textwidth]{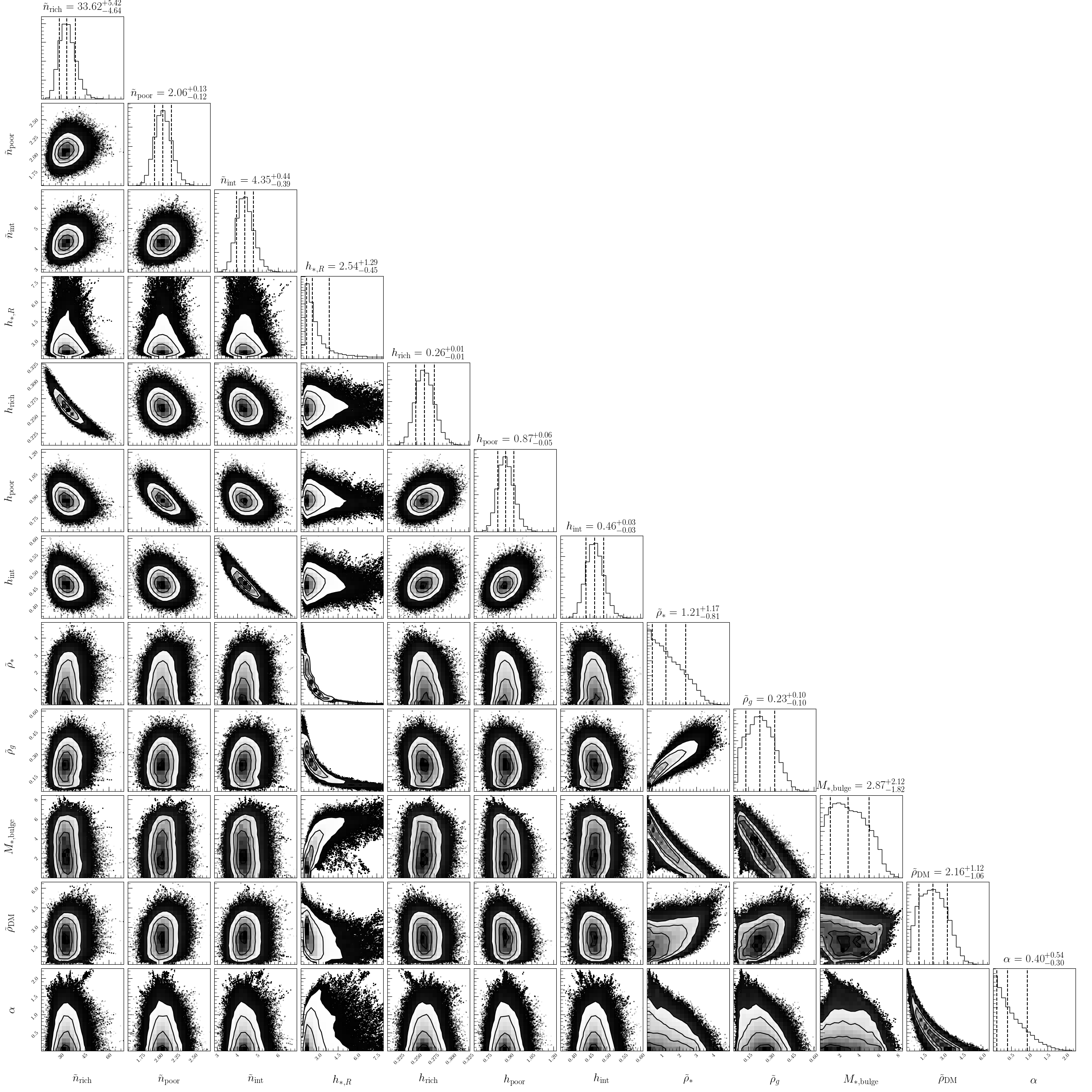}
    \caption{\label{fig:dm_corner} Same as \Fig{fig:mond_corner} except for the dark matter model.}
  \end{center}
\end{figure*}

\begin{figure*}[h]
  \begin{center}
    \includegraphics[width=.45\textwidth]{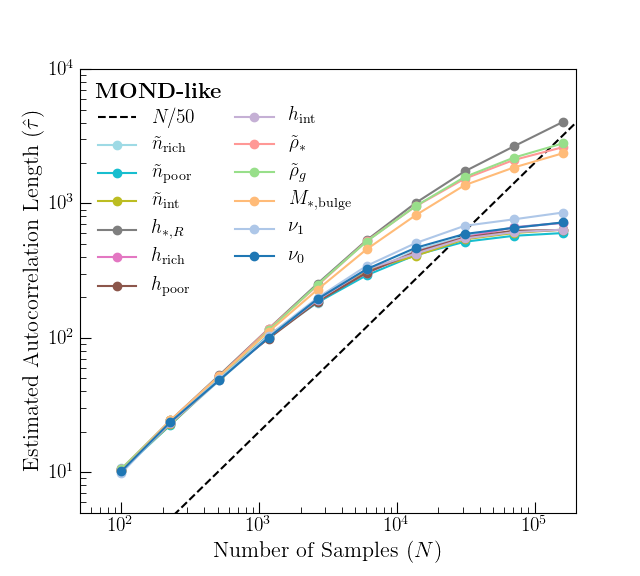}
    \includegraphics[width=.45\textwidth]{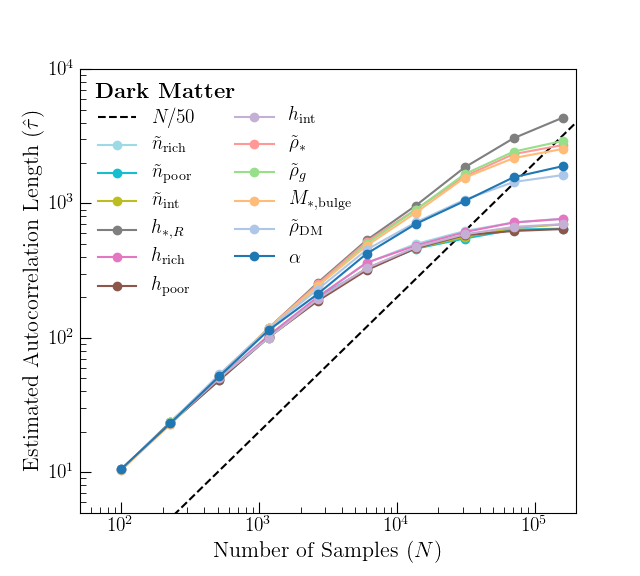}
    \caption{\label{fig:autocorr_plots} Autocorrelation lengths ($\hat{\tau}$) computed as a function of the number of post burn-in MCMC samples for the MOND-like (\textbf{left}) and dark matter (\textbf{right}) analyses. The \textit{black dashed line} representing $\hat{\tau}=N/50$ is shown as a benchmark of convergence.}
  \end{center}
\end{figure*}

We run \texttt{emcee} with 200 walkers for 200,000 steps per walker, discarding the first 40,000 steps as burn-in. Based on visual inspection of the parameter trace plots, we expect this choice of burn-in to be highly conservative. We note that the limiting factor in burn-in time is the parameter $h_{*,R}$, which gradually diverges and eventually saturates its prior range. A corner plot of all the free parameters and their posterior distributions and two-parameter correlations can be seen for the ML model in \Fig{fig:mond_corner} and for the DM model in \Fig{fig:dm_corner}. We assess the convergence of the post burn-in chain by computing the autocorrelation time for each parameter~\cite{2013PASP..125..306F,goodman2010ensemble}. We estimate that the autocorrelation length for each parameter is $\lesssim$~10,000 steps, which suggests that our sample chain is long enough to be a sufficient proxy for the true posterior distribution. In \Fig{fig:autocorr_plots}, we show the estimated autocorrelation length, $\hat{\tau}$, as a function of the number of post burnin-in steps, $N$, as a visual indicator of the convergence for both the ML and DM baseline scans. We repeat similar analyses for the additional \texttt{emcee} scans that we run with additional measurements $h_{*,R,\rm obs}$ and $M_{*,\rm bulge, obs}$ in the likelihood; in general, we find that these scans converge much faster.

\end{appendix}

\clearpage
\def\bibsection{} 
\bibliographystyle{apsrev}
\bibliography{Short_MOND_bib.bib}

\end{document}